\documentclass[letterpaper,12pt]{article}

\pdfoutput=1
\usepackage{amsmath}
\usepackage{amsfonts}
\usepackage{hyperref}
\usepackage{bm}

\makeatletter
\renewcommand\section{\@startsection {section}{1}{\z@}%
{-3.5ex \@plus -1ex \@minus -0.2ex}%
{2.3ex \@plus 0.2ex}%
{\normalfont\normalsize\bfseries}}

\renewcommand\subsection{\@startsection{subsection}{2}{\z@}%
{-3.25ex \@plus -1ex \@minus -0.2ex}%
{1.5ex \@plus 0.2ex}%
{\normalfont\normalsize\bfseries}}

\def\@seccntformat#1{\csname the#1\endcsname.\hspace{0.8em}}
\makeatother

\newcounter{definition}
\newcommand{\definition}[1]{\refstepcounter{definition}
\noindent \textbf{Definition \thedefinition: #1}}

\newcounter{assumption}
\newcommand{\assumption}[1]{\refstepcounter{assumption}
\noindent \textbf{Assumption \theassumption: #1}}

\begin{document}

\setlength{\baselineskip}{3.75ex}
\setlength{\abovedisplayskip}{12pt}
\setlength{\abovedisplayshortskip}{0pt}
\setlength{\belowdisplayskip}{12pt}
\setlength{\belowdisplayshortskip}{12pt}

\noindent
\textbf{\LARGE Probabilistic inference when the population}\\[2ex]
\textbf{\LARGE space is open}\\[3ex]

\noindent
\textbf{Russell J. Bowater}\\
\textit{Independent researcher,
Doblado 110, Col.\ Centro, City of Oaxaca, C.P.\ 68000, Mexico.\\
Email address: as given on arXiv.org. Twitter/X profile:
\href{https://x.com/naked_statist}{@naked\_statist}\\ Personal website:
\href{https://sites.google.com/site/bowaterfospage}{sites.google.com/site/bowaterfospage}}\\[2ex]

\noindent
\textbf{\small Abstract:}
{\small In using observed data to make inferences about a population quantity, it is commonly
assumed that the sampling distribution from which the data were drawn belongs to a given parametric
family of distributions, or at least, a given finite set of such families, i.e.\ the population
space is assumed to be closed.
Here, we address the problem of how to determine an appropriate post-data distribution for a given
population quantity when such an assumption about the underlying sampling distribution is not made,
i.e.\ when the population space is open.
The strategy used to address this problem is based on the fact that even though, due to an open
population space being non-measurable, we are not able to place a post-data distribution over all
the sampling distributions contained in such a population space, it is possible to partition this
type of space into a finite, countable or uncountable number of subsets such that a distribution
can be placed over a variable that simply indicates which of these subsets contains the true
sampling distribution.
Moreover, it is argued that, by using sampling distributions that belong to a number of parametric
families, it is possible to adequately and elegantly represent the sampling distributions that
belong to each of the subsets of such a partition.
Since a statistical model is conceived as being a model of a population space rather than a model
of a sampling distribution, it is also argued that neither the type of models that are put forward
nor the expression of pre-data knowledge via such models can be directly brought into question by
the data.
Finally, the case is made that, as well as not being required in the modelling process that is
proposed, the standard practice of using P values to measure the absolute compatibility of an
individual or family of sampling distributions with observed data is neither meaningful nor
useful.}

\pagebreak

\vspace{4ex}
\section{Introduction}

Conventional theories of statistical inference, e.g.\ the frequentist, Fisherian and Bayesian
approaches to inference, are largely based on the assumption that the true sampling distribution of
the observed data is a member of a given parametric family of distributions, or at the very least,
a member of a given finite set of such families.
In this paper, we will be concerned with addressing the problem of how to make inferences about
quantities of interest on the basis of a given data set in the case where an assumption of this
type about the underlying sampling distribution of the data does not apply, or in other words, in
the case where the population space is open rather than closed.
Before going further, we will clarify, in the section that immediately follows, the terminology
that will be adopted so that we are able to clearly define and then address this important problem.

\vspace{3ex}
\section{Main concepts, terminology and notation}

\vspace{2ex}
\definition{The sampling distribution}

\vspace{1.5ex}
\noindent
Let $p_x(y)$ denote the sampling distribution of the observed data $y=\{y_1,y_2,\ldots,y_n\}$.
This distribution may often be defined so that it depends on additional observable variables $x$,
which will be referred to as conditioning variables, e.g.\ in a regression analysis, these
variables are usually termed `independent variables' or `covariates'.

Observe that, given a particular scheme for probability sampling from the values of $y$ in the
population, the corresponding sampling distribution $p_x(y)$ will completely define this
population, and hence, there is some degree of interchangeability between the terms `sampling
distribution' and `population'.

\pagebreak
\definition{A scientific model}

\vspace{1.5ex}
\noindent
A model will be regarded as being an idealised and approximate description of reality.
Under this definition, the term `true model' is therefore an oxymoron.
To clarify, it can never be known for certain that a model is a perfect description of reality,
since in such circumstances it would cease to be a model.
Also, the term `a set of models' would be inappropriate if we knew for certain that the set must
contain `a model' that perfectly describes reality.

\vspace{5ex}
\definition{A statistical model}

\vspace{1.5ex}
\noindent
A statistical model will be considered to be an approximate description of reality that places
probabilities on the outcomes of uncertain events.
With regard to statistical inference, a type of model that we may often wish to make reference to
is a model of the sampling distribution of the data, i.e.\ an approximate representation of this
sampling distribution.
For example, if we attempt, in some way, to use the observed data $y$ to estimate the sampling
distribution $p_x(y)$ of this data, then this estimate may be regarded as being a statistical
model.

\vspace{5ex}
\definition{The population space}

\vspace{1.5ex}
\noindent
The population space $\mathcal{P}$ is the space of all possible sampling distributions from which
the data $y$ could have been generated given what is known about the true sampling distribution
$p_x(y)$.

\vspace{5ex}
\definition{A closed population space}
\label{def5}

\vspace{1.5ex}
\noindent
A population space will be defined as being closed if it consists of one of the following:

\vspace{1.5ex}
\noindent
1) A given finite set of sampling distributions.

\vspace{1ex}
\noindent
2) A given finite set of parametric families for the sampling distribution of interest $p_x(y)$,
where each family depends on a finite number of parameters and where any given sampling
distribution can not belong to more than one of the parametric families in this~set.

\vspace{1ex}
\noindent
3) A combination of (1) and (2).

\vspace{1.5ex}
If a population space $\mathcal{P}$ is closed, then it will be denoted as the set
$\mathcal{P}_C = \{F_1, F_2,\linebreak \ldots, F_c\}$, where each member $F_i$ of this set is
either an individual sampling distribution or a parametric family of sampling distributions.
Observe that a closed population space is measurable in the sense that it is possible to place a
probability distribution over all the sampling distributions contained in this type of space.

\vspace{5ex}
\definition{An open population space}
\label{def6}

\vspace{1.5ex}
\noindent
If all the sampling distributions contained in a population space can not be contained in a
population space that satisfies the condition in Definition~\ref{def5} to be a closed population
space, then the population space will be described as being an open population space.
Let this type of population space be denoted as the set $\mathcal{P}_O$. An example of an open
population space could simply be the space of all possible sampling distributions that could have
generated the data set $y$ in the case where nothing was known about the true sampling distribution
$p_x(y)$ before this data set was collected. Observe that an open population space is
non-measurable in the sense that it is not possible to place a probability distribution over all
the sampling distributions contained in this type~of~space.

\vspace{5ex}
\definition{A model of the population space}
\label{def7}

\vspace{1.5ex}
\noindent
A model of the population space, which will be denoted as $\mathcal{M}$, will be assumed to be
simply an approximate representation of the genuine population space $\mathcal{P}$.
It will be supposed that this type of model satisfies the condition in Definition~\ref{def5} to be
a closed population space, i.e.\ a population space model $\mathcal{M}$ has to be specified
according to the expression $\mathcal{M}=\{F_1, F_2, \ldots, F_m\}$, where each member $F_i$ of
this set is either an individual sampling distribution or a parametric family of sampling
distributions, and where any given sampling distribution can not belong to more than one of the
parametric families in this set.
However, unlike a closed population space, a population space model $\mathcal{M}$ will not be
assumed to contain a sampling distribution that has the precise form of the true sampling
distribution $p_x(y)$.

Observe that if the set $\mathcal{M}$ is being used to approximate a closed population space
$\mathcal{P}_C$, then the number of elements $m$ in the former set will, in general, be different
from the number of elements $c$ in the latter set. Also, it will usually be inappropriate to
\linebreak regard each element of $\mathcal{M}$ as approximating a corresponding element of
$\mathcal{P}_C$, instead we will generally need to regard the whole of the set $\mathcal{M}$ as
approximating the whole of the~set~$\mathcal{P}_C$.

Let us now digress to discuss in a little more detail how a population space model $\mathcal{M}$
may be regarded as actually modelling an open or closed population space. For a population space
model $\mathcal{M}$ to be useful, it will in general contain far fewer sampling distributions,
either in the form of its individual elements $F_i$ or as members of the parametric families that
it contains, than the population space that it is designed to model.
Therefore, each sampling distribution in the population space model $\mathcal{M}$ needs to be
viewed as representing a small set of sampling distributions in the population space $\mathcal{P}$,
which would be natural to regard as forming a neighbourhood of this space $\mathcal{P}$ around the
sampling distribution concerned from the model $\mathcal{M}$.
In this way, a population space model $\mathcal{M}$ can be viewed as placing pegs in the population
space being modelled, where the pegs are sampling distributions belonging to the model
$\mathcal{M}$ that represent the sampling distributions that lie in the vicinity of the peg.
By pegging out the population space in such a manner, a population space model $\mathcal{M}$ is
able to represent all the possible sampling distributions that lie in a population space
$\mathcal{P}$ independent of whether that population space is closed or open.

To clarify, it should be possible to identify a partition of the population space of interest
$\mathcal{P}$, which we will denote as the partition $S$, that bijectively maps onto the set of all
sampling distributions that are contained in the population space model $\mathcal{M}$, where all
the sampling distributions contained in any given member of this partition can be viewed as being
represented by the sampling distribution in the set $\mathcal{M}$ onto which they map.
If this latter sampling distribution is a member of a parameter family of distributions $F_i$, then
let this individual distribution be denoted as $F_i[\theta^{(i)}]$, where $\theta^{(i)}$ are the
values of the parameters of the family $F_i$ that correspond to the distribution concerned, and let
the member of the partition $S$ of the population space of interest $\mathcal{P}$ that maps onto
the distribution $F_i[\theta^{(i)}]$ be denoted as the set $P_i[\theta^{(i)}]$.
We will refer to the bijective mapping under discussion between the partition $S$ of the population
space $\mathcal{P}$ and all the sampling distributions belonging to the model $\mathcal{M}$ as
being the representation mapping $R$. Finally, let us define the set of sampling distributions
$P_i$ as follows:
\begin{equation}
\label{equ1}
P_i = \bigcup_{\mbox{\footnotesize $\theta^{(i)}\hspace{-0.2em} \in\hspace{-0.1em} \Theta^{(i)}$}}
P_i[\theta^{(i)}]
\vspace{1ex}
\end{equation}
where $\Theta^{(i)}$ is the set of all possible values of $\theta^{(i)}$.
Therefore, the set $P_i$ surjectively maps onto the set of all sampling distributions that belong
to the parametric family $F_i$.

Nevertheless, it will be assumed that, in any given modelling scenario of the type being
considered, it will not be necessary to explicitly identify the partition $S$ of the population
space concerned or the representation mapping $R$. We only need to assume that we have at least a
vague understanding of what could be a suitable general form for the partition $S$ and the mapping
$R$, and also that this partition and this mapping could be genuinely identified if we had the time
to dedicate to that often complex task.

Let us though give an example of how this task could be performed in a relatively simple situation.
In particular, let us imagine that the population space model $\mathcal{M}$ contains four
parametric families $F_1$, $F_2$, $F_3$ and $F_4$ each of which depend on two parameters $\mu$ and
$\sigma$, which are the mean and standard deviation (s.d.) of the sampling distributions concerned.
Furthermore, we will suppose that $F_1$ is a family of left-skewed and leptokurtic (heavy-tailed)
distributions, $F_2$ is a family of left-skewed and platykurtic (light-tailed) distributions, $F_3$
is a family of right-skewed and leptokurtic distributions and $F_4$ is a family of right-skewed and
leptokurtic distributions.
Finally, the population space for this example will be assumed to be the open population space of
all possible sampling distributions for the observed data $y$.

In this example, we could consider specifying the partition $S$ of the open population space of
interest $\mathcal{P}_O$ by partitioning this population space into the following subsets of this
space:
\vspace{0.5ex}
\[
S = \{\hspace{0.1em} P_i[\hspace{0.1em}\mu,\sigma\hspace{0.05em}]: i \in \{1,2,3,4\}, \mu \in
\mathbb{R}\ \mbox{and}\ \sigma \in \mathbb{R}_{>0}\hspace{0.1em} \}
\]
where $P_1[\hspace{0.05em}\mu,\sigma]$ is the set of all sampling distributions with a mean $\mu$
and a s.d.\ $\sigma$ that are left-skewed and leptokurtic, $P_2[\hspace{0.05em}\mu,\sigma]$ is the
set of all sampling distributions with a mean $\mu$ and a s.d.\ $\sigma$ that are left-skewed and
platykurtic, $P_3[\hspace{0.05em}\mu,\sigma]$ is the set of all sampling distributions with a mean
$\mu$ and a s.d.\ $\sigma$ that are right-skewed and leptokurtic, and finally,
$P_4[\hspace{0.05em}\mu,\sigma]$ is the set of all sampling distributions with a mean $\mu$ and a
s.d.\ $\sigma$ that are right-skewed and platykurtic. Specifying the partition $S$ of the
population space $\mathcal{P}_O$ in this way implies, given the assumptions that have been made,
that there is a representation mapping $R$ from the set $S$ to the set:
\[
F = \{\hspace{0.1em} F_i[\hspace{0.1em}\mu,\sigma\hspace{0.05em}]: i \in \{1,2,3,4\}, \mu \in
\mathbb{R}\ \mbox{and}\ \sigma \in \mathbb{R}_{>0}\hspace{0.1em} \}
\]
where, just to clarify, any given element $P_i[\hspace{0.05em}\mu,\sigma]$ of the set $S$ maps onto
the corresponding element $F_i[\hspace{0.05em}\mu,\sigma]$ of the set $F$ and where the sampling
distribution $F_i[\hspace{0.05em}\mu,\sigma]$ is defined according to the definition of the
distribution $F_i[\theta^{(i)}]$ given earlier.
Observe that choosing the representation mapping $R$ to have this particular form would appear to
be quite reasonable since, for example, it is consistent with the left-skewed and leptokurtic
sampling distribution $F_1[\hspace{0.05em}\mu,\sigma]$ being an adequate representation of all the
sampling distributions in the open population space $\mathcal{P}_O$ that are left-skewed and
leptokurtic and that have the same mean and s.d.\ as $F_1[\hspace{0.05em}\mu,\sigma]$, i.e.\ a mean
of $\mu$ and a s.d.\ of $\sigma$.
We have therefore put forward a sensible way of specifying a partition $S$ of the population space
of interest and a representation mapping $R$ in this example.

\vspace{5ex}
\definition{A population quantity}

\vspace{1.5ex}
\noindent
A population quantity $Q$ will be defined as being a characteristic of the true sampling
distribution $p_x(y)$ such as its mean, variance or a given quantile of this distribution.
We may express any given population quantity $Q$ as being the expected value of a given function
$q(y)$ over the sampling distribution $p_x(y)$, i.e.\ the quantity $Q$ may be expressed in the
following way:
\vspace{1ex}
\[
Q = \mbox{E}\hspace{0.1em}[q(y)] = \int_{\mathbb{R}^n} q(y) \pi_x(y) d^{\hspace{0.03em}n}y
\vspace{1ex}
\]
where $\pi_x(y)$ is the density function of the distribution $p_x(y)$.
Also, if the sampling dis\-tri\-bution $p_x(y)$ belongs to a given parametric family of
distributions $F_i$ that depends on a set of parameters $\theta^{(i)} =
\{ \theta^{(i)}_j\hspace{-0.1em}: j=1,2,\ldots,r_i \}$, where $r_i$ is any positive integer, then
clearly a population quantity of interest $Q$ may be expressed as a function of the
parameters~$\theta^{(i)}$.

\pagebreak
\definition{Probabilistic inference}
\label{def9}

\vspace{1.5ex}
\noindent
A method of statistical inference will be classed as being a method of probabilistic inference if
its goal is to place probability distributions over given population quantities of interest $Q$
that represent what is known about these quantities after the data $y$ have been observed.
If, as is usually the case, these quantities of interest $Q$ need to be regarded as being fixed but
unknown values, then clearly post-data probability distributions of these quantities will
inevitably be, in some sense, subjective probability distributions.
However, this does not exclude the possibility that it could be often the case that many would
consider a post-data distribution of this type to be an extremely good and perhaps almost objective
representation of what is known about the population quantity $Q$ concerned after the data $y$ have
been observed.

Two examples of theories of inference that may be applied to obtain post-data distributions of
quantities of interest $Q$ are organic fiducial inference, as developed in
Bowater~(2018, 2019, 2021) and Bayesian inference.
Furthermore, the fiducial-Bayes fusion as put forward in Bowater~(2023) is a general theory of
inference that allows us to obtain post-data distributions of this type by combining organic
fiducial inference and Bayesian inference in a way that respects, where it is appropriate, the
capacity of these two approaches to inference to be used on their own.
Nevertheless, these methods of inference are designed to make post-data inferences about quantities
of interest $Q$ in the case where the population space is closed, and as a consequence, can not be
directly applied to the case where the population space is open, which, as has already been made
clear, will be the case of primary interest in the present paper.

\vspace{5ex}
\assumption{Simplifying the discussion}

\vspace{1.5ex}
\noindent
Without a great loss of generality, it will be assumed that each member $F_i$ of any given closed
population space $\mathcal{P}_C = \{F_1, F_2, \ldots, F_c\}$ or any given population space model
$\mathcal{M}=\{F_1, F_2, \ldots, F_m\}$ that will be considered in this paper is a parametric
family of sampling distributions rather than an individual sampling distribution.
This assumption is being made to simplify the terminology that needs to be used in the discussion
of ideas that will be presented.
Also, from this discussion, the reader should be able to easily appreciate how to go about tackling
the problem of inference that is of interest when this simplifying assumption is not made.

\vspace{3ex}
\section{Aim clarified and overview}

In terms of the definitions given in the previous section, let us clarify that the main aim of the
present paper will be to propose ways in which probabilistic inference, according to
Definition~\ref{def9}, may be adequately performed in the case where the population space is open,
according to Definition~\ref{def6}.

Let us now briefly describe the structure of the remainder of this paper. In the next section, even
though it will not be the main focus of our attention, we discuss the problem of how to make
inferences about a population quantity $Q$ on the basis of a given data set $y$ in the case where
the population space is closed.
The core strategy for addressing this problem in the case where the population space is open is
then laid out in Sections~\ref{sec1} to~\ref{sec4} in a manner that gradually progresses from
outlining the overall principles on which this strategy is based to putting forward ways of dealing
with specific issues regarding the application of this strategy in practice.

In the latter part of the paper, topics associated with how the strategy in question relates to and
contrasts with what is currently viewed as being standard practice in statistical modelling are
discussed.
More specifically, in Section~\ref{sec6}, we consider how much more work would generally need to be
carried out in applying the strategy proposed in Section~\ref{sec5} compared to applying a more
conventional type of modelling strategy.
Follow\-ing on from this, in Section~\ref{sec7}, we address the issue of whether, in using the
approach to statistical modelling put forward in the present paper, P values may be regarded as
having the same kind of important role in assessing the compatibility of a parametric family of
sampling distributions with an observed data set as they would usually be regarded as having in a
standard modelling process.
As a last point of discussion, the issue is considered in Section~\ref{sec8} of whether, or in what
way, it is possible for a statistical model as conceived in the present paper or the expression of
pre-data knowledge through such a model to be brought into question by data that may be collected.

To conclude the paper, we emphasize in Section~\ref{sec9} the key elements of the approach to
inference that the paper has developed and discussed.
We should also point out that, after the bibliography, there is an appendix that gives further
technical details about this approach to inference in the case where the population space is
closed.

\vspace{3ex}
\noindent
\section{Inference when the population space is closed}

With regard to cases in which the population space is closed, i.e.\ cases where the population
space consists of the finite set of distributional families $\mathcal{P}_C = \{F_1, F_2, \ldots,
F_c\}$, it is helpful to imagine that the post-data distribution of any given population quantity
$Q$ is determined by a two-step procedure in which, first, post-data probabilities
$\pi(F_1\,|\,y)$, $\pi(F_2\,|\,y), \ldots, \pi(F_c\,|\,y)$ are derived for the hypotheses that the
true sampling distribution $p_x(y)$ is a member of the parametric families $F_1, F_2, \ldots, F_c$,
respectively, and second, the post-data density function of the quantity $Q$ is determined
conditional on the sampling distribution $p_x(y)$ belonging to each of the families in the set
$\{F_1, F_2, \ldots, F_c\}$, i.e.\ in this second step, the set of conditional post-data density
functions $\{\pi(Q\,|\,F_i,y):i=1,2,\ldots,c\}$ is determined.
Having performed these calculations, the overall post-data density function $\pi(Q\,|\,y)$, i.e.\
the post-data density of the quantity of interest $Q$ unconditional on the sampling distribution
$p_x(y)$ being a member of any particular family $F_i$ in the set
$\mathcal{P}_C=\{F_1, F_2, \ldots, F_c\}$, can then be obtained by using the following expression:
\vspace{0.5ex}
\begin{equation}
\label{equ2}
\pi(Q\,|\,y) = \sum_{i=1}^{c} \pi(Q\,|\,F_i,y) \pi(F_i\,|\,y)
\vspace{0.5ex}
\end{equation}

We may determine the set of post-data probabilities $\{\pi(F_i\,|\,y) : i=1,2,\ldots,c\}$, the set
of conditional post-data density functions $\{\pi(Q\,|\,F_i,y):i=1,2,\ldots,c\}$, and therefore,
the overall post-data density $\pi(Q\,|\,y)$ by using a straightforward application of Bayesian
theory in which prior probabilities $\pi(F_1), \pi(F_2), \ldots, \pi(F_c)$ are assigned to the
hypotheses that the true sampling distribution $p_x(y)$ is a member of the parametric families
$F_1, F_2, \ldots, F_c$, respectively, and in which prior density functions are placed over the
parameters $\theta^{(i)}$ of each of the families $F_i$ conditional on the distribution $p_x(y)$
belonging to the family $F_i$ concerned, i.e.\ prior density functions that are naturally denoted
as the set of functions $\{\pi(\theta^{(i)}\,|\,F_i): i=1,2,\ldots,c\}$.
However, in implementing this Bayesian method, the common practice of choosing the prior densities
of the parameters on which distributional families depend to be improper densities, which, when
applied to the present case, means choosing the prior densities $\{\pi(\theta^{(i)}\,|\,F_i):
i=1,2,\ldots,c\}$ to be improper densities, can lead to a procedure for determining the overall
post-data density $\pi(Q\,|\,y)$ that may not have a useful interpretation even from a mathematical
viewpoint, let alone a philosophical viewpoint (see for example Shafer~1982 and O'Hagan~1995).

On the other hand, in the Appendix of the present paper, it is shown how the method of inference
that is put forward in Bowater~(2022), which belongs to the theory of the fiducial-Bayes fusion
outlined in Bowater~(2023), may be naturally extended to determine the overall post-data density
$\pi(Q\,|\,y)$ in the case where the population space is closed via the two-step procedure
associated with the use of equation~(\ref{equ2}) that has just been described.
This method of inference avoids the aforementioned difficulty in applying the Bayesian approach to
inference to the problem of interest by taking into account pre-data knowledge about the parameters
$\theta^{(i)}$ of any given distributional family $F_i$, or a lack of such knowledge, in a way that
is different to how this type of knowledge is taken into account when using the Bayesian approach.

\vspace{3ex}
\noindent
\section{Inference when the population space is open}
\label{sec5}

\vspace{1.5ex}
\subsection{General strategy}
\label{sec1}

If the population space is open, then clearly we can not apply exactly the same approach to
inference as outlined in the previous section to determine a post-data distribution for a given
quantity of interest $Q$. Moreover, we are faced with the problem of how to determine this
post-data distribution when, due to the population space being non-measurable, we are not able to
place a post-data distribution over all the sampling distributions contained in this space.

Our strategy for resolving this problem will be based on observing that, even though an open
population space is non-measurable, it is possible to partition this type of population space into
various subsets such that a probability distribution can be placed over a variable that is defined
so that it simply indicates which one of these subsets contains the true sampling distribution
$p_x(y)$, and this can be achieved independent of whether the number of subsets that constitute the
partition concerned is assumed to be finite, countable or uncountable.
Also, we may observe that, for each of these subsets, it may be possible to find a sampling
distribution that adequately represents the other sampling distributions in the subset concerned,
even under the restriction that these representative distributions have to belong to some type of
global parametric structure over which a post-data distribution can be easily placed.
Finally, let us point out that, within each of the subsets of the type in question of a given open
population space $\mathcal{P}_O$, the difference between the maximum and minimum likelihood of
observing the data set $y$ may be quite small.

Taking into account these observations, it will be assumed that, in general, the most appropriate
way to derive an overall post-data density for a quantity of interest $Q$, i.e.\ a post-data
density of the type $\pi(Q\,|\,y)$, is by modelling the open population space $\mathcal{P}_O$
concerned. In particular, if we model this population space $\mathcal{P}_O$ by using a model of the
type $\mathcal{M}=\{F_1, F_2, \ldots, F_m\}$ as specified in Definition~\ref{def7}, then we may
obtain a post-data density of the kind $\pi(Q\,|\,x)$ by using equation~(\ref{equ2}), but with the
closed population space $\mathcal{P}_C$ replaced by the population space model $\mathcal{M}$.

In deciding to use this method of inference, we are faced, though, with the following two
fundamental issues:

\vspace{2ex}
\noindent
1) How to choose which and how many parametric families $F_i$ to include in the population space
model $\mathcal{M}=\{F_1, F_2, \ldots, F_m\}$ so that pre-data knowledge about the true sampling
distribution $p_x(y)$ can be expressed adequately.

\vspace{2ex}
\noindent
2) How to adjust for the fact that we will be applying a method of inference that is based on the
assumption that the true sampling distribution $p_x(y)$ belongs to one of the parametric families
of distributions $F_i$ contained in the set $\mathcal{M}=\{F_1, F_2, \ldots, F_m\}$ when in reality
this assumption will be almost certainly false and may not even be close to being true.

\vspace{2ex}
We will address the first of these two issues in the following section.

\pagebreak
\noindent
\subsection{Expressing pre-data knowledge through the population space model}
\label{sec2}

\vspace{2.5ex}
\assumption{General expression of pre-data knowledge}
\label{asm2}

\vspace{1.5ex}
\noindent
It will assumed that to be able to apply the general strategy put forward in the previous section
to determine an overall post-data density for a given population quantity $Q$, i.e.\ a post-data
density of the type $\pi(Q\,|\,y)$, pre-data knowledge about the true sampling distribution
$p_x(y)$ needs to be expressed through the population space model $\mathcal{M}$ at the following
three levels:

\vspace{1.5ex}
\noindent
1) At the level of the specification of the population space model $\mathcal{M}=\{F_1, F_2, \ldots,
F_m\}$, i.e.\ at the level of deciding which distributional families $F_i$ should be included in
this model.

\vspace{1.5ex}
\noindent
2) At the level of deciding, for each value of $i$ in the set $\{1,2,\ldots,m\}$, what pre-data or
prior probability should be assigned to the hypothesis that the sampling distribution $p_x(y)$ lies
in the set $P_i$, where this latter set is as defined in equation~(\ref{equ1}).

\vspace{1.5ex}
\noindent
3) At the level of deciding, for each value of $i$ in the set $\{1,2,\ldots,m\}$, the form of a
particular function of the parameters $\theta^{(i)}$ of the family $F_i$ that represents pre-data
knowledge about the true sampling distribution $p_x(y)$, or a lack of such pre-data knowledge,
conditional on this sampling distribution lying in the region $P_i$ of the population space
concerned.

\vspace{3ex}
Let us now clarify all that has just been assumed and discuss how, in practice, we may go about
expressing pre-data knowledge about the sampling distribution $p_x(y)$ at the three different
levels in question.

We will begin by highlighting that one of the options we clearly have for expressing pre-data
knowledge about $p_x(y)$ at the first level referred to in Assumption~\ref{asm2} is to choose to
include very few distributional families $F_i$ in the population space model $\mathcal{M}$, e.g.\
only one distributional family, and to choose these families to be distributional families that
depend on a very small number of parameters.
However, it will, of course, generally not be possible to optimally express the depth of our
pre-data knowledge about the sampling distribution $p_x(y)$ over such a simple population space
model, and this will often result in the overall post-data density $\pi(Q\,|\,y)$ being an
inadequate representation of what we know about the quantity of interest $Q$ after the data $y$
have been observed, more specifically, it may well be inadequate in the sense of being an overly
confident representation of what we know about this quantity.

At the other extreme, we may decide to do what L.\ J.\ Savage once recommended we should do when
faced with the problem of choosing a parametric family of distributions to model an underlying
sampling distribution of interest and try to make this family of distributions `as big as a house'.
In the present context, this strategy could be interpreted as choosing to include a very large
number of distributional families $F_i$ in the set $\mathcal{M}$ and choosing these families to be
highly flexible distributional families that each depend on a very large number of parameters.
However, in general, it will be very difficult to express our pre-data knowledge about the sampling
distribution $p_x(y)$ over such a com\-plex population space model $\mathcal{M}$, and of course, if
this pre-data knowledge is not taken into account in a sensible manner in determining a post-data
distribution for the quantity of interest $Q$, then the adequacy of this post-data distribution as
a representation of our post-data knowledge about this quantity may be brought into question.

Also, in relation to the issue being discussed, a natural question that arises is, if we are in a
situation where a single parametric family of distributions $F_1$ that depends on a small number of
continuous parameters is not an adequate model for the open population space of interest, is it
best, in general, to try to replace this parametric family with a more flexible parametric family
$F_*$ that depends on a larger number of continuous parameters $\theta^{(*)}$, or incorporate other
parametric families $F_2, F_3, \ldots, F_m$ into the population space model $\mathcal{M}$ that also
depend on a small number of parameters?
Since the former strategy may be considered to be more mathematically elegant than the latter
strategy, many are of the opinion that, in general, it is the most appropriate way in which to
address the issue in question (see for example Draper~1995, 1999).
However, by using the latter strategy, it may well be possible to construct a population space
model $\mathcal{M}$ that represents both the main areas and difficult-to-reach corners of the open
population space of interest better than using the former strategy, while relying on a total number
of parameters, i.e.\ the total number of parameters in the set $\{\theta^{(1)}, \theta^{(2)},
\ldots, \theta^{(m)}\}$, that is less than the number of parameters that would need to be relied on
when the alternative strategy in question is used, i.e.\ the number of parameters in the set
$\theta^{(*)}$.

Furthermore, we may observe that, in using the former strategy, pre-data knowledge about the true
sampling distribution $p_x(y)$ will need to be expressed over all the possible combinations of the
values of the parameters that belong to the entire set $\theta^{(*)}$, whereas in using the latter
strategy, this pre-data knowledge will need to be expressed simply over all the possible
combinations of the values of the parameters that belong to each of the individual elements
$\theta^{(i)}$ of the set $\{\theta^{(1)}, \theta^{(2)}, \ldots, \theta^{(m)}\}$.
Therefore, even if the total number of parameters in the set $\{\theta^{(1)}, \theta^{(2)}, \ldots,
\theta^{(m)}\}$ is greater than the number of parameters in the set $\theta^{(*)}$, it may well be
far less complicated to express our pre-data knowledge about the parameters that belong to the set
$\{\theta^{(1)}, \theta^{(2)}, \ldots, \theta^{(m)}\}$ rather than the parameters that belong to
the set $\theta^{(*)}$, or to put it another way, it may well be far easier to express our pre-data
knowledge about the sampling distribution $p_x(y)$ when using the latter strategy that is under
discussion rather than the former strategy.

With regard to the expression of pre-data knowledge about the sampling distribution $p_x(y)$ at the
second level referred to in Assumption~\ref{asm2}, let us imagine that we are at the point of
introducing one or more distributional families into the population space model
$\mathcal{M}=\{F_1, F_2, \ldots, F_m\}$ that represent regions of the open population space of
interest $\mathcal{P}_O$ that overlap with a region of this space that is represented by a
distributional family $F_j$ that is already included in the model $\mathcal{M}$.
Of course, on introducing the new distributional families of the type in question into the model
$\mathcal{M}$, it is likely to be considered appropriate that the region $P_j$ of the population
space $\mathcal{P}_O$ that is represented by the distributional family $F_j$ is reduced in size,
which will generally imply that the pre-data or prior probability of the sampling distribution
$p_x(y)$ lying in the updated set $P_j$ will also need to be reduced.
In other words, if we introduce one or more distributional families into the population space model
$\mathcal{M}$ that are similar to a distributional family $F_j$ that is already included in this
model, then it is likely that the prior probability of the sampling distribution $p_x(y)$ lying in
the region of the population space $\mathcal{P}_O$ that is represented by the family $F_j$ will
need to be reduced.

Finally, let us discuss how we may go about expressing pre-data knowledge about the sampling
distribution $p_x(y)$ at the third level referred to in Assumption~\ref{asm2}. To begin with, we
will assume that the function of the parameters $\theta^{(i)}$ that is used, under
Assumption~\ref{asm2}, to express pre-data knowledge about the sampling distribution $p_x(y)$
conditional on $p_x(y)$ lying in the set $P_i$ is a function that, for all possible values of the
parameters $\theta^{(i)}$, assigns a probability density value to the hypothesis that the sampling
distribution $p_x(y)$ lies in the set $P_i[\theta^{(i)}]$ conditional on $p_x(y)$ lying in the set
$P_i$, where the set $P_i[\theta^{(i)}]$ is as specified in Definition~\ref{def7}.
Observe that this assumption implicitly asserts that a probability density value rather than a
probability mass should be assigned to a hypothesis that states that the sampling distribution
$p_x(y)$ lies in a given set of the type $P_i[\theta^{(i)}]$.
Before going further, let us clarify why this assertion is indeed valid.

In doing this, it will be assumed, without a great loss of generality, that the set of parameters
$\theta^{(i)}$ consists of a single univariate parameter~$\theta$ and the space of this parameter
is the real line.
Having made this assumption, let us proceed by defining $A$ to be a countable partition of the
space of the parameter $\theta$ such that each member of this partition can be expressed as an
interval of the form $[\theta_0, \theta_0 + \delta\theta_0]$, where $\theta_0$ is a given value of
$\theta$ and $\delta\theta_0$ is a small positive value that may depend on the value of $\theta_0$.
Also, we will define the set of sampling distributions $P_i[\theta_0, \theta_0 + \delta\theta_0]$
as follows:
\vspace{0.5ex}
\[
P_i[\theta_0, \theta_0 + \delta\theta_0] = \bigcup_{\mbox{\footnotesize $\theta
\in [\theta_0, \theta_0\hspace{-0.05em} +\hspace{-0.05em}\delta\theta_0]$}} P_i[\theta]
\vspace{0.5ex}
\]

In relation to these definitions, it may be pointed out that, it is meaningful to consider
assigning a positive probability to every hypothesis in the countable set of hypotheses that
consists of the hypothesis that $p_x(y)$ lies in the set $P_i[\theta_0, \theta_0 + \delta\theta_0]$
for every interval $[\theta_0, \theta_0 + \delta\theta_0]$ in the partition $A$.
Expanding on this point, let us suppose that, for any given interval
$[\theta_0, \theta_0 + \delta\theta_0]$ in the partition $A$, the probability being referred to,
i.e.\ the probability $\pi(\hspace{0.05em}p_x(y) \in P_i[\theta_0, \theta_0 + \delta\theta_0])$,
can be defined as follows:
\vspace{1ex}
\[
\pi(\hspace{0.1em}p_x(y) \in P_i[\theta_0, \theta_0 + \delta\theta_0]) =
\int_{\mbox{\footnotesize $\theta_0$}}^{\mbox{\footnotesize \hspace{0.1em}$\theta_0\hspace{-0.05em}
+\hspace{-0.05em}\delta\theta_0$}} h(\theta)d\theta
\vspace{1ex}
\]
where $h(\theta)$ is a given continuous density function of $\theta$.
Under this assumption, it is clear that as $\delta\theta_0$ tends to zero, the probability
$\pi(\hspace{0.05em}p_x(y) \in P_i[\theta_0, \theta_0 + \delta\theta_0])$ will tend to the value
$h(\theta_0)\delta\theta_0$. Therefore, it is meaningful to refer to a probability density value of
$h(\theta)$ being assigned to the hypothesis that the sampling distribution $p_x(y)$ lies in any
given set of the type $P_i[\theta]$.

Nevertheless, conditional on the sampling distribution $p_x(y)$ lying in any given region $P_i$ of
an open population space of interest $\mathcal{P}_O$, the task of assigning a probability density
value to the hypothesis of $p_x(y)$ lying in the set $P_i[\theta^{(i)}]$ for all possible values of
the parameters $\theta^{(i)}$ is generally going to be extremely complicated.
The reason for this is that not only, under the assumptions being made, will there be an
uncountable number of sets of the type $P_i[\theta^{(i)}]$ contained in the region $P_i$, but also
that, as already mentioned, it will be very difficult, in many practical situations, to precisely
specify the set $P_i[\theta^{(i)}]$ for all values of the parameters~$\theta^{(i)}$.

As a direct consequence of this in the particular case being discussed and for related reasons in
other cases, it will be assumed that the function of the parameters $\theta^{(i)}$ that is used,
under Assumption~\ref{asm2}, to express pre-data knowledge about the sampling distribution $p_x(y)$
conditional on $p_x(y)$ lying in the set $P_i$ is determined in the way that is set out in the
following assumption.

\vspace{3ex}
\assumption{Expressing pre-data knowledge at the lowest level}
\label{asm3}

\vspace{1.5ex}
\noindent
It will be assumed that, for each value of $i$ in the set $\{1,2,\ldots,m\}$, the function of the
parameters $\theta^{(i)}$ of the family $F_i$ that, according to Assumption~\ref{asm2}, is used to
represent pre-data knowledge about the sampling distribution $p_x(y)$, or a lack of such pre-data
knowledge, conditional on this sampling distribution lying in the region $P_i$ is determined by
imagining that the distribution $p_x(y)$ must be a member of the distributional family $F_i$, i.e.\
for all values of the parameters $\theta^{(i)}$, it is imagined that if $p_x(y)$ lies in the set
$P_i[\theta^{(i)}]$, then it must be equal to the distribution $F_i[\theta^{(i)}]$.

\vspace{3ex}
As an alternative to making Assumption~\ref{asm3}, we could consider making this assumption just to
obtain a rough approximation to the function of the parameters $\theta^{(i)}$ that is used to
represent pre-data knowledge about $p_x(y)$ conditional on $p_x(y)$ lying in the region $P_i$, and
then try to adjust this approximation on the basis of the way in which the distribution
$F_i[\theta^{(i)}]$ represents the distributions in the set $P_i[\theta^{(i)}]$ over all possible
values of the parameters $\theta^{(i)}$ with the aim of obtaining a more appropriate version of the
function in question.
We could imagine applying this strategy when the function of the parameters $\theta^{(i)}$ being
referred to is, for example, a probability density function of the parameters $\theta^{(i)}$, which
of course would be the case when using the Bayesian approach to inference, or when it is, for
example, a global pre-data (GPD) function of the parameters $\theta^{(i)}$ as defined in
Bowater~(2019, 2021), which may be assumed to be the case when using the theory of organic fiducial
inference outlined in these earlier papers.

However, for reasons already given, making the kind of adjustment being considered to an
approximate form of a function for representing pre-data knowledge about $p_x(y)$ of the type that
is under discussion is, in general, going to be an extremely complicated task.
Also, we should point out that, in many practical situations, determining this function of the
parameters $\theta^{(i)}$ using Assumption~\ref{asm3} without any subsequent adjustment to this
function being made is arguably going to be a very adequate strategy.
Therefore, it is not very often going to be both viable and appropriate to use the more complicated
strategy just mentioned.

\vspace{3ex}
\noindent
\subsection{Clarification of the general strategy}
\label{sec3}

Before going further, let us clarify what will be the proposed general approach to inference in the
case where the population space is open.

\vspace{3ex}
\assumption{Determining the post-data density of a population quantity}

\vspace{1.5ex}
\noindent
It will be assumed that an appropriate post-data density of any given population quantity $Q$,
i.e.\ a post-data density of the type $\pi(Q\,|\,y)$, may be derived by proceeding via the
following three steps:

\vspace{1.5ex}
\noindent
1) We determine, for each value of $i$ in the set $\{1,2,\ldots,m\}$, what post-data probability
should be assigned to the hypothesis that the true sampling distribution $p_x(y)$ lies in the
region $P_i$ of the open population space of interest. Let this set of post-data probabilities be
denoted as $\{\pi(P_i\,|\,y): i=1,2,\ldots,m\}$.

\vspace{1.5ex}
\noindent
2) We determine, for each value of $i$ in the set $\{1,2,\ldots,m\}$, a post-data probability
density function for the parameters $\theta^{(i)}$ conditional on the sampling distribution
$p_x(y)$ being a member of the parametric family of distributions $F_i$. Let this set of post-data
density functions be denoted as $\{\pi(\theta^{(i)}\,|\,F_i,y): i=1,2,\ldots,m\}$.

\vspace{1.5ex}
\noindent
3) We calculate the post-data density $\pi(Q\,|\,y)$ by using the following formula:
\vspace{0.5ex}
\begin{equation}
\label{equ3}
\pi(Q\,|\,y) = \sum_{i=1}^{m} \int_{\mbox{\footnotesize $\Theta^{(i)}$}}
\pi(Q\,|\,F_i[\theta^{(i)}])\pi(\theta^{(i)}\,|\,F_i,y) \pi(P_i\,|\,y) d\theta^{(i)}
\vspace{1ex}
\end{equation}
where $\pi(Q\,|\,F_i[\theta^{(i)}])$ is the sampling density function of the population quantity
$Q$ given that $F_i[\theta^{(i)}]$ is the sampling distribution of the data in question and where
again $\Theta^{(i)}$ is the set of all possible values of the parameters $\theta^{(i)}$.

\vspace{3ex}
Observe that the presence of the density function $\pi(\theta^{(i)}\,|\,F_i,y)$ in
equation~(\ref{equ3}) rather than the more general density function
$\pi(P_i[\theta^{(i)}]\,|\,P_i,y)$, i.e.\ the post-data density that $p_x(y)$ lies in the set
$P_i[\theta^{(i)}]$ conditional on the sampling distribution $p_x(y)$ lying in the region $P_i$, is
a consequence of it being implicitly assumed that, in the construction of this latter post-data
density, if the sampling distribution $p_x(y)$ lies in any given set of the type
$P_i[\theta^{(i)}]$, then it must be equal to the distribution $F_i[\theta^{(i)}]$, i.e.\ we are
implicitly adhering to the assumption that underlies the method outlined in Assumption~\ref{asm3}.

In the next two sections, we will take a look at strategies for performing the first and second
steps of the procedure just outlined and the impact of these strategies on the third step of this
procedure, i.e.\ the calculation of the post-data density $\pi(Q\,|\,y)$ using the formula given in
equation~(\ref{equ3}).

\vspace{3ex}
\noindent
\subsection{Determining post-data probabilities of \texorpdfstring{$p_x(y)$}{p.x(y)} lying in the
regions \texorpdfstring{$P_i$}{P.i}}

In this section, we will consider the problem of determining the post-data probabilities
$\{\pi(P_i\,|\,y): i=1,2,\ldots,m\}$, which is a task that we are required to perform in the first
step of the general approach for deriving the post-data density $\pi(Q\,|\,y)$ outlined in the
previous section.

Let us begin by pointing out that a possible strategy for resolving this problem may be arrived at
by simply making the assumption that, over all values of $i$ in the set $\{1,2,\ldots,m\}$, the
sampling distribution $p_x(y)$ must belong to the parametric family of distributions $F_i$ if it
belongs to the region $P_i$ of the open population space of interest.
Under this assumption, the problem of determining the set of probabilities $\{\pi(P_i\,|\,y):
i=1,2,\ldots,m\}$ reduces, of course, to the problem of determining the set of probabilities
$\{\pi(F_i\,|\,y): i=1,2,\ldots,m\}$.
To determine this latter set of post-data probabilities, we may use either the Bayesian approach to
inference or the method of inference outlined in Bowater~(2022), the application of which to the
case of interest is clarified in the Appendix of the present paper.
Observe that, if standard Bayesian theory is applied to the case of interest, then the
probabilities $\{\pi(F_i\,|\,y): i=1,2,\ldots,m\}$ will, in effect, be determined by what is known
as Bayesian model averaging, which is an inferential framework that is discussed, for example, in
Draper~(1995), Clyde~(1999) and Hoeting \textit{et al.}~(1999).

However, a major drawback of using the general strategy that has just been proposed is that, in
determining the set of probabilities $\{\pi(P_i\,|\,y): i=1,2,\ldots,m\}$, it does not, of course,
take into account that the sampling distributions belonging to any given parametric family $F_i$
represent all of the sampling distributions belonging to the corresponding region $P_i$ of the open
population space of interest.
For example, a given parametric family $F_i$ may be assigned a relatively low post-data probability
$\pi(F_i\,|\,y)$ when using this type of strategy due to the fact that the sampling distributions
which it contains poorly account for outlying values in the data set $y$, and yet in the region
$P_i$ that this family represents in the population space, there may be sampling distributions to
which these outlying values fit very adequately.

In trying to address the overall issue that is being referred to, it may be helpful if we focus our
attention on the following general question:

\vspace{2ex}
\noindent
Given two regions $P_i$ and $P_j$ of the open population space of interest $P_{O}$, i.e.\ the
regions represented by the distributional families $F_i$ and $F_j$, respectively, how should we go
about determining the post-data probability that the true sampling distribution $p_x(y)$ lies in
the region $P_i$, i.e.\ the probability $\pi(P_i\,|\,y)$, relative to the post-data probability
that $p_x(y)$ lies in the region $P_j$, i.e.\ the probability $\pi(P_j\,|\,y)$?

\vspace{2ex}
Observe that, due to the regions $P_i$ and $P_j$ being non-measurable, we will not be able to place
probability distributions over all the sampling distributions contained in these two regions, and
therefore, it will not be possible to address this question using standard Bayesian inference or
the method of inference outlined in Bowater~(2022).
Also, although we could consider extending the theory of organic fiducial inference put forward in
Bowater~(2019, 2021) to cope with the fact that the regions $P_i$ and $P_j$ are non-measurable by
using ideas related to the concept of `approximation regions' discussed in, for example,
Davies~(2014, 2018), it would appear difficult to justify the benefits that could be obtained by
following such a path.
For these reasons, a method of inference will now be considered for addressing the issue under
discussion that does not naturally belong to the same class of methods as the approaches to
inference just highlighted, i.e.\ a method of inference that does not naturally form part of the
theory of the fiducial-Bayes fusion outlined in Bowater~(2023).

To begin with, let us imagine that we attempt to loosely identify a set
$P_{i*} = \{P_{ik}: k=1,2,\ldots,a\}$ of all those sub-regions of the region $P_i$ of the open
population space of interest $P_{O}$ that possess the following two properties:

\vspace{1.5ex}
\noindent
1) Each of these sub-regions had a substantial pre-data or prior probability of containing the true
sampling distribution $p_x(y)$.

\vspace{1.5ex}
\noindent
2) The sampling distributions that each of these sub-regions contain all give a substantial
probability or probability density to the event of generating a data set that is equal to the data
set $y$ that was actually observed.

\vspace{1.5ex}
\noindent
Also, we will imagine that we attempt to loosely identify a set
$P_{j*} = \{P_{jk}: k=1,2,\linebreak \ldots,b\}$ of all those sub-regions of the region $P_j$ of
the population space $P_{O}$ that possess the same two properties just put forward.
Obviously, since the adjective `substantial' that appears in the definitions of the two properties
in question is open to interpretation, the definitions of the sets
$P_{i*} = \{P_{ik}: k=1,2,\ldots,a\}$ and $P_{j*} = \{P_{jk}: k=1,2,\ldots,b\}$ will neither be
unique nor precise, but nevertheless the way that these two sets are being defined is adequate for
achieving the goal that is of current interest.

Having loosely identified the two sets of sub-regions $P_{i*}$ and $P_{j*}$, we will next assume
that we attempt to assign a value to the following probability ratio:
\vspace{1ex}
\begin{equation}
\label{equ4}
\frac{\pi(P_i\,|\,y)}{\pi(P_j\,|\,y)}
\vspace{1ex}
\end{equation}
where both the post-data probabilities in this ratio are as previously defined, in a way that
adequately takes into account the following relevant quantities:

\vspace{1.5ex}
\noindent
1) The pre-data probabilities of the true sampling distribution $p_x(y)$ lying in each of the
sub-regions in the sets $P_{i*} = \{P_{ik}: k=1,2,\ldots,a\}$ and
$P_{j*} = \{P_{jk}: k=1,2,\ldots,b\}$.

\vspace{1.5ex}
\noindent
2) For each of the $a+b$ sub-regions in the two sets $P_{i*}$ and $P_{j*}$, the minimum probability
or probability density of generating a data set that is equal to the data set $y$ over the sampling
distributions in the sub-region concerned.

\vspace{1.5ex}
Although the assignment of a value to the probability ratio in equation~(\ref{equ4}) on the basis
of the quantities just referred to depends on a judgement with a large subjective element, this
element can be regarded as being much smaller than the subjective element on which the elicitation
of the pre-data probabilities $\pi(P_i)$ and $\pi(P_j)$ depends, i.e.\ the pre-data probabilities
that $p_x(y)$ lies in the regions $P_i$ and $P_j$, respectively, since it is a judgement that can
be based on concrete empirical data, i.e.\ the data set $y$.
Furthermore, in many situations, the possibility that the sampling distribution $p_x(y)$ lies in
the region $P_i$ or the region $P_j$ will be virtually eliminated by observing the data $y$, and
therefore, in these situations, one or both of the post-data probabilities $\pi(P_i\,|\,y)$ and
$\pi(P_j\,|\,y)$ could be reasonably judged as being close to zero.
By contrast, another type of situation that is also likely to often arise is where the regions
$P_i$ and $P_j$ lie so close to each other in the population space of interest $\mathcal{P}_O$ that
there exists very little opportunity for the data set $y$ to provide empirical grounds for choosing
the ratio of post-data probabilities $\pi(P_i\,|\,y) / \pi(P_j\,|\,y)$ to be much different from
the ratio of pre-data probabilities $\pi(P_i) / \pi(P_j)$. In such cases, it may be quite
reasonable to try to avoid excessive subjectivity in the assignment of a value to the ratio of
post-data probabilities in equation~(\ref{equ4}) by applying the following rule:
\vspace{1.5ex}
\[
\frac{\pi(P_i\,|\,y)}{\pi(P_j\,|\,y)} = \frac{\pi(P_i)}{\pi(P_j)}
\vspace{1ex}
\]

Clearly, if it is possible to determine ratios of post-data probabilities of the type given in
equation~(\ref{equ4}) by using the general kind of strategy that has just been put forward, then it
will be possible to determine the set of post-data probabilities
$\{\pi(P_i\,|\,y): i=1,2,\ldots,m\}$. In particular, if we are able to determine the ratio of
post-data probabilities in equation~(\ref{equ4}) for any given fixed value of $j$ and for all the
values of $i$ in the set $\{1,2,\ldots,m\}$, then the set of post-data probabilities
$\{\pi(P_i\,|\,y): i=1,2,\ldots,m\}$ may be determined via the normalisation of the probability
ratios in question, provided that $\pi(P_j\,|\,y)>0$.

Observe that, for each of the sub-regions that are members of the set
$P_{i*} = \{P_{ik}: k=1,2,\ldots,a\}$ or the set $P_{j*} = \{P_{jk}: k=1,2,\ldots,b\}$, it will
become, in general, more difficult to assign, even in a loose way, a pre-data probability to the
hypothesis that the true sampling distribution $p_x(y)$ lies in the sub-region concerned as the
sizes of these two sets of sub-regions become larger, i.e.\ as the values $a$ and $b$ increase.
Also, it is clear that, as the sizes of the sets $P_{i*}$ and $P_{j*}$ become larger, the larger
the number of relevant quantities that will need to be taken into account in applying the strategy
for assessing the ratio of post-data probabilities in equation~(\ref{equ4}) that has just been
outlined, and as a consequence, the more complicated it will become to implement this strategy.
We may conclude, therefore, that the strategy in question will be more useful if some kind of steps
are taken to prevent the sizes $a$ and $b$ of the sets $P_{i*}$ and $P_{j*}$ from becoming
extremely large.

\vspace{3ex}
\noindent
\subsection{Determining post-data densities of the parameters
\texorpdfstring{$\theta^{(i)}$}{theta'(i)} of the families \texorpdfstring{$F_i$}{F.i}}
\label{sec4}

For each value of $i$ in the set $\{1,2,\ldots,m\}$, an appropriate post-data density function of
the parameters $\theta^{(i)}$ conditional on the true sampling distribution $p_x(y)$ being a member
of the parametric family of distributions $F_i$, i.e.\ a post-data density of the type
$\pi(\theta^{(i)}\,|\,F_i,y)$, may be sensibly determined by using a straightforward application of
either the subjective Bayesian approach to inference or the theory of the fiducial-Bayes fusion as
outlined in Bowater~(2023), which, as already mentioned, includes subjective Bayesian inference as
a special case.
We may observe that expressing pre-data knowledge about the sampling distribution $p_x(y)$, or a
lack of such pre-data knowledge, in accordance with Assumption~\ref{asm3} presented earlier,
facilitates the use of these approaches to inference in constructing the type of post-data density
function in question.
However, given that, in the general theory of inference being put forward in the present paper,
each sampling distribution $F_i[\theta^{(i)}]$ in any given distributional family $F_i$ is assumed
to represent all the sampling distributions in the region $P_i[\theta^{(i)}]$ of the open
population space of interest $\mathcal{P}_O$, we may ask how closely, in any given situation, the
post-data density function $\pi(\theta^{(i)}\,|\,F_i,y)$ would approximate the post-data density
function $\pi(P_i[\theta^{(i)}]\,|\,P_i,y)$, where this latter density function is as defined in
Section~\ref{sec3}\hspace{0.1em}?

To answer this question, we should first take into account the justification that was given in
Section~\ref{sec2} for expressing pre-data knowledge about the true sampling distribution $p_x(y)$,
or a lack of such pre-data knowledge, in accordance with Assumption~\ref{asm3}.
Having done this, we then need to consider the consequences of constructing the post-data density
$\pi(P_i[\theta^{(i)}]\,|\,P_i,y)$ under the condition that the sampling distribution $p_x(y)$ is
not only contained in the region $P_i$ of the population space $\mathcal{P}_O$, but is also a
member of the distributional family $F_i$ when in fact this latter assumption may not be valid.
Although there is no doubt that, in various situations, the effect of making this false assumption
will not be negligible, it will unfortunately be very complicated, in general, to try to measure
the size of this effect.
The good news, though, is that in most practical situations, instead of needing to be able to
exactly measure the size of the effect in question, we will only need to justify why there should
be a high degree of confidence that, when accumulated over all values of $i$ in the set
$\{1,2,\ldots,m\}$, the size of this effect will not be large enough to imply that the post-data
density of the population quantity of interest $Q$ defined by equation~(\ref{equ3}) will be an
inadequate representation of what we know about this quantity having observed the data set $y$.

For this reason, let us now make the case that, for any given value of $i$ in the set
$\{1,2,\ldots,m\}$, the practical suitability of using the post-data density function
\hspace{0.05em}$\pi(\theta^{(i)}\,|\,F_i,y)$ as an approximation to the post-data density function
$\pi(P_i[\theta^{(i)}]\,|\,P_i,y)$ may be adequately checked by comparing the post-data density of
the quantity of interest $Q$ conditional on the sampling distribution $p_x(y)$ being a member of
the distributional family $F_i$, i.e.\ the post-data density $\pi(Q\,|\,F_i,y)$, with the post-data
density of $Q$ conditional on $p_x(y)$ being a member of the distributional family $F^*_i$, i.e.\
the post-data density $\pi(Q\,|\,F^*_i,y)$, where the family $F^*_i$ can be viewed as representing
the same region $P_i$ of the population space that is represented by the family $F_i$.
More specifically, it will be assumed that the distributional family $F^*_i$ is chosen to be as
different from the distributional family $F_i$ as possible under the condition that the sampling
distributions that belong to the family $F^*_i$ must be considered as representing the sampling
distributions that are contained in the region $P_i$ just as well as the sampling distributions
that belong to the family $F_i$, which is a condition that, of course, constrains the family
$F^*_i$ to being, in general, not that dissimilar from the family $F_i$.
Let us clarify that it is being assumed that the post-data density $\pi(Q\,|\,F_i,y)$ and the
post-data density $\pi(Q\,|\,F^*_i,y)$ are derived by applying the theory of the fiducial-Bayes
fusion with pre-data knowledge about the true sampling distribution $p_x(y)$ being expressed as
best as possible under the condition that $p_x(y)$ is a member of the family $F_i$ in the first
case and that $p_x(y)$ is a member of the family $F^*_i$ in the second case.

If, in using the strategy just referred to, it turns out that the post-data density
$\pi(Q\,|\,F^*_i,y)$ closely approximates the post-data density $\pi(Q\,|\,F_i,y)$, then we may
well feel that we are reasonably justified in regarding the post-data density function
$\pi(\theta^{(i)}\,|\,F_i,y)$ as being an adequate approximation to the post-data density function
$\pi(P_i[\theta^{(i)}]\,|\,P_i,y)$ for the purpose of constructing a post-data density of the
quantity of interest $Q$ conditional on the sampling distribution $p_x(y)$ lying in the region
$P_i$ of the given population space $\mathcal{P}_O$, i.e.\ the post-data density
$\pi(Q\,|\,P_i,y)$.
On the other hand, if in using this strategy, it turns out that the post-data density
$\pi(Q\,|\,F^*_i,y)$ is a poor approximation to the post-data density $\pi(Q\,|\,F_i,y)$, then we
may consider including an additional distributional family $F_{+}$ in the population space model
$\mathcal{M}=\{F_1, F_2, \ldots, F_m\}$ that is similar to the distributional family $F_i$.
This additional parametric family of distributions may be, for example, the distributional family
$F^*_i$.

In doing this, the region of the population space that is represented by the distributional family
$F_i$ will, in general, naturally become smaller since part of the region of this space that was
represented by the family $F_i$, i.e.\ the original region $P_i$, will now be represented by the
new distributional family that has been included in the model $\mathcal{M}$, i.e.\ the family
$F_{+}$.
As a result, the distributional family $F_i$ should be able, in general, to represent the region
$P_i$ as defined in relation to the new population space model $\mathcal{M}_{+}$ better than it
represented the region $P_i$ when this region was defined in relation to the old population space
model $\mathcal{M}$.
Furthermore, this means that if the strategy that has just been outlined is repeated with respect
to the new population space model $\mathcal{M}_{+}$, then the distributional family $F^*_i$, which
in repeating this strategy would need to be redefined so that it represents, in an alternative
manner, the sampling distributions in the new region $P_i$ of the population space, would be, in
general, more similar to the distributional family $F_i$ than was the case when the strategy in
question was implemented with respect to the original population space model $\mathcal{M}$.
Finally, as a consequence of this latter observation, the post-data density $\pi(Q\,|\,F^*_i,y)$
defined with respect to the respecified distributional family $F^*_i$ may well be considered to be
a much better approximation to the post-data density $\pi(Q\,|\,F_i,y)$ than when it was defined
with respect to the original distributional family $F^*_i$ to the extent that we may now be willing
to regard the post-data density function $\pi(\theta^{(i)}\,|\,F_i,y)$ as being an adequate
approximation to the post-data density function $\pi(P_i[\theta^{(i)}]\,|\,P_i,y)$ for the purpose
of constructing the post-data density $\pi(Q\,|\,P_i,y)$, where the region $P_i$ is now defined in
relation to the new population space model $\mathcal{M}_{+}$.

If this is not considered to be the case, then we may consider enlarging the population space model
$\mathcal{M}$ once again so that it contains yet another or various more distributional families
that are similar to the distributional family $F_i$. The strategy under discussion may therefore
repeat itself once more, and if necessary, keep repeating itself in this way until it is considered
that the post-data density function $\pi(\theta^{(i)}\,|\,F_i,y)$ is an adequate approximation to
the post-data density function $\pi(P_i[\theta^{(i)}]\,|\,P_i,y)$ for achieving the goal of current
interest.

To illustrate how the strategy that has just been put forward could be applied, let us suppose that
the region $P_i$ of the open population space of interest $\mathcal{P}_O$ contains only
right-skewed sampling distributions and that these distributions are equally well represented by
the family of log-normal distributions, which will be defined to be the distributional family
$F_i$, and the family of gamma distributions, which will be defined to be the distributional family
$F^*_i$.
Since any given log-normal distribution and any given gamma distribution is completely specified by
its mean $\mu$ and its variance $\sigma^2$, let us, in this particular scenario, make the further
assumption that, for all values of $\mu$ and $\sigma^2$, the right-skewed distributions in the set
$P_i[\hspace{0.05em}\mu,\sigma^2]$, where this set is defined as the set $P_i[\theta^{(i)}]$ was
defined in Definition~\ref{def7}, are equally well represented by the log-normal distribution
$F_i[\hspace{0.05em}\mu,\sigma^2]$ and the gamma distribution $F^*_i[\hspace{0.05em}\mu,\sigma^2]$,
i.e.\ the log-normal and gamma distributions that have a mean equal to the value $\mu$ and a
variance equal to the value $\sigma^2$.

In following the strategy in question, we now need to compare the post-data density of a population
quantity of interest $Q$ conditional on the true sampling distribution $p_x(y)$ being a log-normal
distribution, i.e.\ the post-data density $\pi(Q\,|\,F_i,y)$, with the post-data density of the
quantity $Q$ conditional on $p_x(y)$ being a gamma distribution, i.e.\ the post-data density
$\pi(Q\,|\,F^*_i,y)$.
To give an example, if the quantity of interest $Q$ was the population mean $\mu$, then it would be
expected that, due to the central limit theorem, the post-data density $\pi(Q\,|\,F^*_i,y)$ would
be regarded as being an adequate approximation to the post-data density $\pi(Q\,|\,F_i,y)$ at least
for a data set $y$ of a moderate size. Furthermore, if the quantity of interest $Q$ was, even say,
the population variance $\sigma^2$, it may well be the case that, in the example under discussion,
the post-data density $\pi(Q\,|\,F^*_i,y)$ would be considered to adequately approximate the
post-data density $\pi(Q\,|\,F_i,y)$.
On the other hand, if the quantity of interest $Q$ was a quantile in the tails of the sampling
distribution $p_x(y)$, then it may well be the case that the post-data density $\pi(Q\,|\,F^*_i,y)$
would be viewed as being a poor approximation to the post-data density $\pi(Q\,|\,F_i,y)$.

In this latter case, we may of course continue to apply the strategy that was outlined earlier by
enlarging the population space model $\mathcal{M}$ so that it contains an additional distributional
family $F_{+}$ that is similar to the log-normal family of distributions $F_i$, e.g.\ the gamma
family of distributions $F^*_i$.
We would then need to assess whether, in this new context, we now would be prepared to regard the
post-data density function $\pi(\theta^{(i)}\,|\,F_i,y)$ as being an adequate approximation to the
post-data density function $\pi(P_i[\theta^{(i)}]\,|\,P_i,y)$ for the purpose of constructing the
post-data density $\pi(Q\,|\,P_i,y)$. If this is not considered to be the case, then we would have
the option of repeating the exercise in question once more and so on.

Observe that, since both the log-normal and gamma family of distributions can be defined so that
they have the mean $\mu$ and the variance $\sigma^2$ as their two parameters, the adequacy of the
post-data density function $\pi(\theta^{(i)}\,|\,F_i,y)$ as an approximation to the post-data
density function $\pi(P_i[\theta^{(i)}]\,|\,P_i,y)$ for achieving the goal under discussion may
also be assessed, in this particular example, by comparing, for all values of $\mu$ and $\sigma^2$,
the height of the joint post-density $\pi(\mu,\sigma^2\,|\,F_i,y)$, i.e.\ the joint post-density of
$\mu$ and $\sigma^2$ conditional on the true sampling distribution $p_x(y)$ being a member of the
family $F_i$, with the height of the joint post-density $\pi(\mu,\sigma^2\,|\,F^*_i,y)$.
Clearly, if the post-data density $\pi(\mu,\sigma^2\,|\,F^*_i,y)$ closely approximates the
post-data density $\pi(\mu,\sigma^2\,|\,F_i,y)$ for all values of $\mu$ and $\sigma^2$, then this
latter density function may well be considered to be an adequate approximation to the post-data
density function $\pi(P_i[\hspace{0.05em}\mu,\sigma^2]\,|\,P_i,y)$ for the purpose of determining
the post-data density $\pi(Q\,|\,P_i,y)$, where, as we know, the set
$P_i[\hspace{0.05em}\mu,\sigma^2]$ represents all sampling distributions in the region $P_i$ of the
given population space $\mathcal{P}_O$ that have a mean equal to the value $\mu$ and a variance
equal to the value~$\sigma^2$.

\vspace{3ex}
\noindent
\subsection{Treating a closed population space as though it is open}

Let us imagine that we are trying to make post-data inferences about a population quantity of
interest $Q$ in a situation where the true sampling distribution $p_x(y)$ may lie anywhere in a
given open population space $P_{O}$, i.e.\ we are not able to exclude the possibility that this
distribution lies in any positive measure subset of this population space, when we are informed, in
some way, that in fact, the distribution $p_x(y)$ must be a member of a distributional family $F_i$
that, although being unknown, has to belong to a given set $\{F_1, F_2, \ldots, F_c\}$ of
distributional families that forms a subset of the population space $P_{O}$.
At first glance, it may appear that receiving the information in question must always be good news,
since as a consequence, a problem of making post-data inferences about the quantity $Q$ under the
assumption that the population space is open is reduced to a seemingly simpler problem that is
based around a population space that is now closed.
However, since the number of distributional families $F_i$ in the closed population space
$\mathcal{P}_C = \{F_1, F_2, \ldots, F_c\}$ may be very large or this population space may contain
some distributional families that depend on a very large number of parameters, it may be very
difficult, or even impossible, to express our pre-data knowledge about the sampling distribution
$p_x(y)$ by assigning pre-data probabilities to this distribution belonging to each of the
distributional families in the population space $\mathcal{P}_C$, and by placing either prior
density functions or global pre-data (GPD) functions over the parameters on which these families
depend.

For this reason, it may be convenient, in certain cases, to try to model a closed population space
in the same way in which, as was shown in previous sections, an open population space may be
modelled, and therefore, in effect, treat a closed population space as though it is an open
population space.
The application of this strategy will clearly have a good chance of being considered successful if
an adequate model $\mathcal{M} = \{F_1, F_2, \ldots, F_m\}$ can be found for the closed population
space of interest $\mathcal{P}_C$ that contains far fewer distributional families $F_i$ than this
population space, i.e.\ $m$ is far less than $c$, and contains distributional families that, as a
group, depend on far fewer parameters than on which, as a group, the distributional families
contained in the population space $\mathcal{P}_C$ depend.

\vspace{3ex}
\noindent
\section{Corner cutting in the modelling process}
\label{sec6}

In applying the theory that has been outlined in the preceding sections to address the problem of
making post-data inferences about population quantities of interest $Q$ in the case where the
population space is open, it is clear that generally a great deal more work will be required than
would be required if we tried, in some way, to tackle this problem by following current standard
statistical practice, i.e.\ the practice of only fitting one parametric family of sampling
distributions to the data set of interest. However, if the alternative to carrying out all this
extra work would be that the post-data inferences made about the quantities $Q$ would be completely
untrustworthy, then clearly there is no alternative to doing the extra work that is required in
applying the theory of inference in question, i.e.\ cutting corners is not an option.
On the other hand, if we were never allowed to cut corners in performing the task being referred
to, then we would be led to the devastating conclusion that many current model-based methods of
statistical inference are at best inappropriate or at worse completely invalid.
Therefore, we may ask in what circumstances would it perhaps be acceptable to cut corners in the
modelling process in the way that is being referred to, and in what circumstances this practice
could definitely be regarded as being foolhardy.

Under the assumption that the population space is open, making post-data inferences about
quantities of interest $Q$ on the basis of a single parametric family of sampling distributions is
likely to be acceptable in the case where the open population space concerned can be regarded as
being `not that far away' from being a closed population space that only contains the parametric
family in question.
We may well find ourselves in this type of situation when, for example, we are considering whether
to make post-data inferences about a population proportion on the basis of a population space model
$\mathcal{M}$ that only contains the binomial family of distributions, or when we are considering
whether to make post-data inferences about a population event rate on the basis of a population
space model $\mathcal{M}$ that only contains the Poisson family of distributions.
In either of these cases, if the assumptions that underlie the use of the single parametric family
of sampling distributions concerned have a very low probability of falling down heavily, then we
may well decide that we have an adequate justification for making post-data inferences about the
parameter of interest simply on the basis of this single distributional family, rather than
carrying out all the extra work that would be involved in trying to model the open population space
concerned using a variety of distributional families in accordance with the theory of inference
that was outlined in previous sections.
Nevertheless, in following this time-saving course of action, it would be sensible, of course, to
bear in mind that inferences made about the parameter of interest may be, to some extent, overly
precise.

Another way in which we may be able to cut corners in the modelling process of an open population
space $\mathcal{P}_O$ arises due to the possible insensitivity of post-data inferences about a
quantity of interest $Q$ to how we choose to express our pre-data knowledge about the true sampling
distribution $p_x(y)$ through the population space model $\mathcal{M}$.
In accordance with what was discussed in Section~\ref{sec2}, let us assume that in attempting to
adequately express our pre-data knowledge about the sampling distribution $p_x(y)$ by carefully
choosing which distributional families $F_i$ should be included in the population space model
$\mathcal{M}$, we eventually decide that this model is most suitably specified as
$\mathcal{M}=\{F_1, F_2, \ldots, F_m\}$, i.e.\ as a set of $m$ given distributional families.
Next, having determined the post-data densities of the quantity of interest $Q$ conditional on the
sampling distribution $p_x(y)$ being a member of each of the distributional families in the set
$\mathcal{M}$, i.e.\ having determined the set of post-data densities $\{\pi(Q\,|\,F_i,y):
i=1,2,\ldots,m\}$, let it be imagined that this set of post-data densities may be partitioned into
$k$ subsets, where $k<m$, such that, for each of these subsets, either only one post-data density
$\pi(Q\,|\,F_i,y)$ is contained in the subset or the post-data densities $\pi(Q\,|\,F_i,y)$ that
are contained in the subset closely approximate each other.
Finally, we will suppose that from each of these subsets, one post-data density $\pi(Q\,|\,F_i,y)$
is chosen and that the set $\mathcal{M}^* = \{F^*_1, F^*_2, \ldots, F^*_k\}$ is the set of
distributional families $F_i$ on which these selected post-data densities are conditioned.

Clearly, if the overall post-data density of the population quantity $Q$, i.e.\ the post-data
density $\pi(Q\,|\,y)$, was derived on the basis of the population space model $\mathcal{M}^*$
rather than the model $\mathcal{M}$, then this post-data density would be expected to be, in
general, a good approximation to the form that this post-data density would take if it was derived
on the basis of the model $\mathcal{M}$ as long as the post-data probabilities of the true sampling
distribution $p_x(y)$ lying in each of the regions $P^*_1, P^*_2, \ldots, P^*_k$, i.e.\ the
post-data probabilities $\pi(P^*_1\,|\,y), \pi(P^*_2\,|\,y), \ldots, \pi(P^*_k\,|\,y)$, are
determined in a way that is in accordance with how the post-data probabilities $\pi(P_1\,|\,y),
\pi(P_2\,|\,y), \ldots, \pi(P_m\,|\,y)$ would have been determined.
Of course, to be able to convert the procedure just outlined into a useful way of cutting corners
in the modelling process of a given open population space $\mathcal{P}_O$, we would have to possess
the foresight to be able to specify the set of distributional families
$\mathcal{M}^* = \{F^*_1, F^*_2, \ldots, F^*_k\}$ without needing to determine the set of post-data
densities $\{\pi(Q\,|\,F_i,y): i=1,2,\ldots,m\}$, and better still, without even needing to
explicitly specify the set of distributional families $\mathcal{M}=\{F_1, F_2, \ldots, F_m\}$.
We may be able to gain this foresight through knowledge acquired concerning the characteristics of
the distributional families that are included, or we could choose to include, in the set
$\mathcal{M}=\{F_1, F_2, \ldots, F_m\}$. This knowledge may be acquired through analytic results,
simulation studies or practical experience.

To give an example, let us suppose that the data values $y$ are measured on a continuous univariate
scale and that little was known about the true sampling distribution $p_x(y)$ before the data set
$y$ was observed. Under these assumptions, a population space model $\mathcal{M}$ through which our
pre-data knowledge about the distribution $p_x(y)$ could be adequately expressed would naturally
contain a wide variety of distributional families $F_i$, some of which may, for example, be
families of symmetric distributions, families of left-skewed distributions, families of
right-skewed distributions, families of heavy-tailed distributions, etc.
However, if the quantity of interest $Q$ is the population mean $\mu$ and the data set $y$ is of at
least moderate size, then, due to the central limit theorem, it may well be the case that, for
every distributional family $F_i$ that we may imagine could be included in an appropriate model
$\mathcal{M}$ of the open population space of interest $\mathcal{P}_O$, the post-data density
$\pi(Q\,|\,F_i,y)$ would be closely approximated by the post-data density of $Q$ conditional on the
sampling distribution $p_x(y)$ belonging to the normal family of distributions, which we will
denote as the post-data density $\pi(Q\,|\,\mbox{N},y)$.
Therefore, in this case, it would seem reasonable to define the set of distributional families
$\mathcal{M}^*$ such that it only contains the normal family of sampling distributions.

It is clear that needing to derive only the post-data density $\pi(Q\,|\,\mbox{N},y)$ rather than
needing to derive the post-data density $\pi(Q\,|\,F_i,y)$ for every parametric family $F_i$ in the
model $\mathcal{M}$ means that, in applying the general corner-cutting strategy that was just put
forward to the example in question, an enormous amount of time can be saved in the construction of
a suitable overall post-data density of the quantity $Q$ in comparison to directly applying the
theory of inference outlined in previous sections.
Nevertheless, in other examples, it would not be generally expected that the time that would be
saved in applying the corner-cutting strategy under discussion would be of a similar magnitude,
assuming of course that applying such a strategy would lead to any kind of time saving at all.
Therefore, the general usefulness of this corner-cutting strategy should be regarded as being an
open issue.

\vspace{3ex}
\noindent
\section{Do P values have a role in the modelling process?}
\label{sec7}

Since P values are often used in practice to check the fit of a parametric family of sampling
distributions to a set of data, this section will be dedicated to determining whether using P
values for this purpose can or should have any role within the approach to statistical inference
that has been put forward in the present paper.

We will assume that using P values to check the fit of a given parametric family of sampling
distributions to a data set of interest is synonymous with checking whether at least one of these
sampling distributions is compatible with the observed data or checking whether the sampling
distributions in the given parametric family that are most supported by the observed data are
compatible with these data. Therefore, to broaden the discussion, let us try to clarify the meaning
of the concept of compatibility in the context of interest.

To begin with, let us divide the concept of compatibility in question into two separate types of
compatibility, namely relative compatibility and absolute compatibility.
Relative compatibility will be assumed to be a way of measuring the compatibility of a given
parametric family $F_i$ of sampling distributions belonging to a given closed population space
$\mathcal{P}_C = \{F_1, F_2, \ldots, F_c\}$ with an observed data set $y$ that is relative to how
well the other parametric families that are contained in this space fit the data set $y$.
In terms of the theory of inference outlined in this paper, relative compatibility is quite
naturally measured by the post-data probabilities that are determined for the true sampling
distribution $p_x(y)$ belonging to each of the distributional families $F_i$ in the closed
population space $\mathcal{P}_C$, i.e.\ the post-data probabilities $\pi(F_1\,|\,y)$,
$\pi(F_2\,|\,y), \ldots, \pi(F_c\,|\,y)$. To clarify, a distributional family $F_i$ would be
regarded as showing a high relative compatibility with the data $y$ if there is a relatively high
post-data probability that it contains the sampling distribution $p_x(y)$ compared to the other
probabilities in the set $\pi(F_1\,|\,y)$, $\pi(F_2\,|\,y), \ldots, \pi(F_c\,|\,y)$, and as showing
a low relative compatibility with the data if there is a relatively low post-data probability that
it contains the distribution $p_x(y)$.

On the other hand, absolute compatibility will be assumed to be a way of measuring the
compatibility of a given parametric family $F_0$ of sampling distributions that belongs to a given
open or closed population space with an observed data set $y$ that is not relative to how well the
sampling distributions in the population space concerned that do not belong to the family $F_0$ fit
the data set $y$.
Therefore, using a given measure of this type of compatibility, we would be able to divide the
infinite set of all parametric families of distributions that belong to an open population space
$\mathcal{P}_O$ into two groups with one group containing the parametric families that are most
compatible with the data set $y$ of interest and the other group containing the parametric families
that are least compatible with this data set.
However, an important question is whether the compatibility of a family of sampling distributions
with a given data set can, in fact, be adequately measured in this way, i.e.\ whether absolute
compatibility can genuinely be regarded as being both a meaningful and useful concept.

To try to answer this question, it is convenient to examine how the compatibility of a family of
sampling distributions with an observed data set differs as a concept from the plausibility that a
given family of sampling distributions contains the true sampling distribution $p_x(y)$ assessed on
the basis of having observed a given data set $y$.
In this respect, let us begin by comparing this concept of plausibility with the concept of
relative compatibility that was defined earlier. In doing this, it quickly becomes apparent that
there is very little or no meaningful difference between the two concepts being compared.
This is because the plausibility of any given hypothesis conditional on having observed a given
data set is usually measured by the post-data probability of this hypothesis, and as mentioned
earlier, post-data probabilities are natural measures of the relative compatibilities of families
of sampling distributions with an observed data set.

Before advancing this discussion further by comparing the concept of plausibility in the context
being referred to with the concept of absolute compatibility that was just defined, let us pause to
consider how we may actually go about measuring absolute compatibility.

The most serious, and arguably the only serious, proposal that has been made for measuring absolute
compatibility is based on the use of P values.
In particular, for an appropriately chosen test statistic, we may consider measuring this type of
compatibility using the P value that corresponds to the null hypothesis that the true sampling
distribution $p_x(y)$ is equal to the sampling distribution in the distributional family of
interest $F_0$ that best fits the observed data $y$. Alternatively, we may consider measuring
absolute compatibility using the following weighted average of P values:
\vspace{1ex}
\begin{equation}
\label{equ5}
\lambda = \left. \int_{\Theta} \alpha(\theta) w(\theta) d\theta \middle/ \int_{\Theta} w(\theta)
d\theta \right.
\end{equation}
where:\\[1ex]
(i) The function $\alpha(\theta)$ is the P\hspace{0.4em}value that, for an appropriately chosen
test statistic, would apply in the case where the null hypothesis is the hypothesis that the
sampling distribution $p_x(y)$ is equal to the distribution in the distributional family of
interest $F_0$ that corresponds to the parameters of this family taking the values $\theta$.
Let us denote this latter distribution as $F_0[\theta]$.\\[1ex]
(ii) The function $w(\theta)$ measures the plausibility of the sampling distribution $p_x(y)$ being
equal to the distribution $F_0[\theta]$ that was just defined conditional on $p_x(y)$ belonging to
the distributional family $F_0$. For example, the function $w(\theta)$ may be chosen to be the
function $\pi(\theta\,|\,F_0,y)$, i.e.\ the post-data probability density of the parameters
$\theta$ conditional on the distribution $p_x(y)$ belonging to the distributional family $F_0$.
Observe that if this post-data probability density is derived using the Bayesian method of
inference, then we would usually refer to the quantity $\lambda$ defined by equation~(\ref{equ5})
as the posterior predictive P value, see Rubin~(1984).\\[1ex]
(iii) The set $\Theta$ is the set of all possible values of the parameters $\theta$.

\vspace{1ex}
Let us now evaluate how well these ways of using P values to measure the absolute compatibility of
a hypothesis with a given data set perform in comparison to measuring the plausibility of a
hypothesis after observing a set of data using the post-data probability of this hypothesis.
In particular, let us begin with a simple example.

To construct this example, it will be assumed that there was a certain level of belief before a
data value $y$ was observed that a parameter $\theta$ on which the sampling distribution of $y$
depends would equal zero, but apart from this there were no strong pre-data beliefs about the
parameter $\theta$.
More specifically, let the sampling distribution of the data value $y$ be a uniform distribution
with a mean of $\theta$ and a known variance. We will now imagine that the value $y$ is observed,
and taking this value to be the test statistic, the two-sided P value is calculated for the null
hypothesis that $\theta=0$. In doing this, let us suppose that the P\hspace{0.4em}value in
question turns out to be extremely small, say less than 0.000001. On the basis of this information,
we may ask what inferences could be drawn about the true value of $\theta$ in terms of the concepts
of plausibility and absolute compatibility under discussion?

In this example, the P value indicates that the null hypothesis that $\theta=0$ is extremely
incompatible with the data value $y$. However, the height of the sampling density function at the
observed value of $y$ under this null hypothesis is equal to the maximum height of this sampling
density, and so there would appear to be little reason to have less belief in the hypothesis that
$\theta$ equals zero after observing the data value $y$ than the belief we had in this hypothesis
before this data value was observed.
In fact, if we try to address the problem of inference in question by combining organic fiducial
inference with Bayesian inference in the way that was outlined in Bowater~(2022), then by making a
very natural interpretation of what has been assumed regarding our pre-data beliefs about the
parameter $\theta$, it turns out that the post-data probability we should assign to the null
hypothesis that $\theta=0$ should be equal to whatever would have been the pre-data probability
that we would have assigned to this hypothesis.

In summary, even though the null hypothesis that $\theta=0$ has an absolute compatibility with the
data value $y$ that is extremely low, observing this data value should arguably not diminish in any
way the degree of belief that we should have in this null hypothesis being true.
Since, in contrast to the idea of using P values to measure absolute compatibility, the method of
inference being advocated here to determine the post-data probability that $\theta$ equals zero has
a sound philosophical justification, we may therefore conclude that trying to formalise the concept
of absolute compatibility in the way that is being discussed is neither meaningful nor useful in
the example of interest.

A criticism that may be raised against the example that has just been highlighted is that we should
not only consider the case in which it is assumed that there was certain level of belief, before
the data value $y$ was observed, that the parameter $\theta$ was equal to zero, but also the case
in which this assumption is not made. Indeed, various authors have explored the idea of using P
values to measure the absolute compatibility of a point null hypothesis with an observed set of
data without making the assumption that any level of belief was placed in the null hypothesis of
interest being true before the data were observed, see for example, Davies~(2014, 2018) and
Greenland~(2023).
However, if on observing a data set, it is considered as being impossible that a given null
hypothesis about the population of interest is true, then how could we regard the null hypothesis
as being anything other than 100\% incompatible with the data, unless of course we wish to deprive
the word `compatible' as having any worthwhile meaning?
Therefore, the idea that a P value may be used to measure the absolute compatibility of a null
hypothesis with an observed data set without simultaneously taking into account whether the null
hypothesis had either a zero or a positive probability of being true before the data were observed
must be regarded as having a clear logical flaw.
It is perhaps a measure of the inconvenience of this flaw that the aforementioned authors are
prepared to overlook it so blatantly.

Returning to the main topic of discussion, let us also point out that, apart from the example that
has just been examined, there are many other examples where the idea of using P values to measure
the absolute compatibility of a hypothesis with a given data set clearly lies in strong conflict
with a sensible assessment of the plausibility of the given hypothesis after observing the data set
concerned.
For instance, in a case where a data value $y$ is observed that lies directly under the top of a
tall hill in the extreme tails of the sampling density of $y$ that corresponds to a given point
null hypothesis being true, then according to the P\hspace{0.4em}value, the absolute compatibility
of this hypothesis with the data value $y$ may be extremely low, but the plausibility of this
hypothesis conditional on having observed the data value $y$ may be considered to be much greater
than the plausibility of this hypothesis before this data value was observed.
On the other hand, in a case where the observed data value $y$ lies directly under the bottom of a
deep valley in the main body of the sampling density of $y$ that corresponds to a given point null
hypothesis being true, then according to the P value, the absolute compatibility of this hypothesis
with the data value $y$ may be very high, but the plausibility of this hypothesis conditional on
having observed the data value $y$ may be considered to be much less than the plausibility of this
hypothesis before this data value was observed.

Moreover, with regard to the use of P values not only in specific situations of the type just
mentioned, but in any given situation where the population space is closed, e.g.\ their use in
standard problems of inference in which the sampling distribution is known to be a normal
distribution, it may be demonstrated that the conventional interpretations of a
P\hspace{0.4em}value as a measure of the absolute compatibility of a point null hypothesis with a
given data set will generally not be at all in accordance with the plausibility of this hypothesis
after the data set has been observed as measured by the post-data probability that would be
assigned to this hypothesis by applying the method of inference outlined in Bowater~(2022).

It is accepted, though, by many statisticians, but perhaps not the majority of statisticians, that
using P values to measure the absolute compatibility of a given hypothesis with an observed data
set in situations where the population space is closed is an inadequate and unnecessary practice.
For example, advocates of the Bayesian approach to inference would naturally favour placing a
posterior distribution over all the sampling distributions contained within a given closed
population space $\mathcal{P}_C$ by conditioning on the observed data set in the way that is
prescribed by Bayesian theory, rather than using P values to measure the absolute compatibility of
each of these sampling distributions with the data set concerned.
By contrast, in situations where the population space is open, it would appear that most currently
practising statisticians, including many advocates of the Bayesian approach to inference, would
endorse the use of P values to measure the absolute compatibility of a given parametric family of
sampling distributions with a data set of interest.
Nevertheless, similar to the case where the population space is closed, the practice of using P
values to measure absolute compatibility in the case where the population space is open is based on
a vague intuition rather than on a solid philosophical foundation.
In fact, this practice may be criticised in the case where the population space is open in a
similar way to how we have just criticised this practice in the case where the population space is
closed.

To expand on this point, let us begin by remembering that, even though we are not able to place a
probability distribution over all the sampling distributions that belong to an open population
space $\mathcal{P}_O$, we may partition this type of population space in a way that allows a
post-data probability or probability density to be determined for the hypothesis that the true
sampling distribution $p_x(y)$ belongs to any given member of this partition.
More specifically, the approach to statistical inference outlined in the earlier sections of the
present paper allows post-data probabilities and probability densities of this type to be
determined by bijectively mapping all of the sampling distributions that belong to a model
$\mathcal{M} = \{F_1,F_2,\ldots,F_m\}$ of the open population space $\mathcal{P}_O$ onto a suitably
chosen partition $S$ of this population space, where any given sampling distribution belonging to
the model $\mathcal{M}$ can be considered to represent all of the sampling distributions in the
region of the space $\mathcal{P}_O$ to which it is linked by the mapping in question, which, as
earlier, will be referred to as the representation mapping $R$.

This means that, for any given sampling distribution $g_x(y)$ that belongs to the population space
model $\mathcal{M}$, we may compare the P value that, in relation to an appropriately chosen test
statistic, corresponds to the null hypothesis that the true sampling distribution $p_x(y)$ is equal
to the sampling distribution $g_x(y)$ with the post-data probability or probability density that
the true sampling distribution $p_x(y)$ belongs to the region of the open population space
$\mathcal{P}_O$ that is linked to the sampling distribution $g_x(y)$ by the representation mapping
$R$ just mentioned.
In making comparisons of this kind, it would seem reasonable to conjecture that the conventional
interpretations of the type of P value in question as a measure of the absolute compatibility of
the given sampling distribution $g_x(y)$ with the data set of interest would generally not be at
all in accordance with the post-data probability or probability density that the true sampling
distribution $p_x(y)$ belongs to the region of the population space $\mathcal{P}_O$ represented by
the sampling distribution~$g_x(y)$.
Therefore, we may draw essentially the same conclusion about the adequacy of using P values as a
measure of the absolute compatibility of a sampling distribution with an observed data set in the
case where the population space is open as when the population space is closed, i.e.\ in both these
cases, the use of P values as a measure of the type of compatibility in question would appear to be
neither meaningful nor useful.

Of course, we are not arguing that P values in their own right have no use in statistical
inference, since clearly they convey information about the observed data which is of a kind that
generally will be useful in some way.
For example, they are often an important component in the construction of fiducial intervals for
parameters of interest according to the theory of organic fiducial inference outlined in
Bowater~(2019, 2021).
Nevertheless, given that the only widely accepted way of measuring the absolute compatibility of a
given hypothesis with a data set of interest is through the use of P values, we may also conclude
that the usefulness of this concept of compatibility in statistical inference can be seriously
brought into question.

To bring this section to a close, let us, though, try to take a step back from all the analysis
that has just been presented and ask ourselves whether, in a situation where it would seem
immediately apparent that an observed data set $y$ was almost certainly not generated by a given
sampling distribution $g_x(y)$, we could claim that this sampling distribution is not compatible
with the data set $y$ without trying to take into account the forms of other possible sampling
distributions that could have generated this data set.
In contemplating our answer to this question, we may get a strong intuitive feeling that, without
needing to take into account these other possible sampling distributions, we could indeed claim
that the sampling distribution $g_x(y)$ is not compatible with the data set $y$.
Nevertheless, what has been discussed in the present section demonstrates that we would not be
truly claiming that the sampling distribution $g_x(y)$ has a low absolute compatibility with the
data set $y$, but rather we would either be implicitly claiming that, as a result of observing this
data set, it is implausible that the sampling distribution $g_x(y)$ is, in some sense, `similar' to
the true sampling distribution $p_x(y)$, or if a substantial pre-data probability had been
associated with the possibility that the distribution $g_x(y)$ was equal to the distribution
$p_x(y)$, then we could be implicitly claiming that, as a result of observing the data set $y$, the
probability that the sampling distribution $g_x(y)$ is equal to the true sampling distribution
$p_x(y)$ is negligible.
Therefore, if we were to claim that the sampling distribution $g_x(y)$ is not to any real extent
compatible with the observed data set $y$, then our claim would be implicitly based on the concept
of plausibility or relative compatibility as defined earlier, rather than on the concept of
absolute compatibility.

\vspace{3ex}
\noindent
\section{Expressing pre-data knowledge after seeing the data}
\label{sec8}

Similar to when attempting to carry out statistical inference in cases where the population space
is closed, it is clearly acceptable, when applying the theory of inference outlined in the present
paper, to express pre-data knowledge about the true sampling distribution $p_x(y)$ after we have
already observed the data set $y$ as long as this task is carried out in a way that reflects what
was honestly known about the distribution $p_x(y)$ before this data set was observed.
Furthermore, expressing pre-data knowledge about the sampling distribution\hspace{0.05em} $p_x(y)$
after rather than before the data set $y$ has been observed may allow this task to be adequately
performed in a way that may save a substantial amount of~time.

The reason for this is that, before the data set $y$ is observed, our knowledge about the sampling
distribution $p_x(y)$ may well be very vague, i.e.\ we may well have a great deal of uncertainty
about where in the closed or open population space of interest $\mathcal{P}$ this distribution will
lie, and it is likely that, as a result, we will need to include many distributional families $F_i$
in a suitable model $\mathcal{M} = \{F_1,F_2,\ldots,F_m\}$ of this population space, i.e.\ it is
likely that $m$ will be large.
A considerable amount of effort will, therefore, need to be invested in deciding which
distributional families $F_i$ ought to be included in the population space model $\mathcal{M}$, and
in expressing pre-data knowledge about the sampling distribution $p_x(y)$ over the parameters
$\theta^{(i)}$ of each of the distributional families $F_i$ in this model, as well as in
determining the pre-data probabilities that the distribution $p_x(y)$ will lie in each of the
regions $\{P_1,P_2,\ldots,P_m\}$ of the population space concerned, each of these regions
corresponding, of course, to a distributional family $F_i$ in the model $\mathcal{M}$.

However, on observing the data $y$, it is likely to become implausible that the true sampling
distribution $p_x(y)$ will lie in a large region $A$ of the population space $\mathcal{P}$ that
would have been considered as being a plausible region for this distribution to lie before
observing the data.
For this reason, although, in expressing pre-data knowledge about the sampling distribution
$p_x(y)$, we would, of course, need to respect the fact that, before observing the data $y$, there
was a substantial degree of belief that the distribution $p_x(y)$ would lie in this region of the
population space, we may give much less attention to how we exactly express our pre-data knowledge
about $p_x(y)$ over this region, e.g.\ we may decide to include far fewer distributional families
$F_i$ in the population space model $\mathcal{M}$ that are aimed at representing  our pre-data
knowledge about $p_x(y)$ over the region $A$.
To clarify, this is because the post-data density of any given population quantity $Q$, i.e.\ the
density $\pi(Q\,|\,y)$, is likely to be insensitive to how pre-data knowledge about the sampling
distribution $p_x(y)$ is expressed over the region $A$.
Therefore, by taking advantage of this property, it may be possible to save a substantial amount of
time in expressing pre-data knowledge about the distribution $p_x(y)$ over the entire population
space $\mathcal{P}$ in comparison with having attempted to perform this task before the data set
$y$ was observed.

To establish where the region $A$ just referred to may lie in a population space of interest
$\mathcal{P}$ and the exact form that this region may take, we may choose to make use of
conventional computational and graphical methods for model checking. Therefore, in deciding to use
the theory of inference put forward in the present paper, it is certainly not the case that these
traditional techniques of exploratory data analysis are made completely redundant.
However, since a P value is not, in general, a direct measure of the plausibility of the null
hypothesis on which its calculation is based, methods of the type being referred to that are aimed
at testing a hypothesis about the true sampling distribution $p_x(y)$ by using P values can not be
considered as being the most suitable methods to use in carrying out the kind of analysis in
question.

With regard to the topic that has just been discussed, we may ask, as a final point of discussion,
whether it is possible for a statistical model, as conceived in the present paper, to be brought
into question through the collection of empirical data, or more precisely, whether it is possible
that the choice of a model $\mathcal{M}$ for a population space of interest $\mathcal{P}$ and the
expression of pre-data knowledge about the sampling distribution $p_x(y)$ through this model can be
brought into question in this way.

In answering this question, we first need to point out that if, through the use of the population
space model $\mathcal{M}$, a value of zero is assigned to the pre-data probability that the data
set $y$ will belong to a given positive measure subset of the sampling space, and yet, on observing
this data set, it turns out that it is indeed a member of this subset, then there is clearly a
conflict between the way that pre-data knowledge about the sampling distribution $p_x(y)$ has been
expressed through the model $\mathcal{M}$ and the data set $y$.
However, if it is assumed that it would never be sensible to assign a pre-data probability of zero
to the possibility of observing a data set $y$ that belongs to a given positive measure subset of
the sampling space when this possibility was not ruled out through the definition of the population
space of interest $\mathcal{P}$, then we may argue that the conflict in question is the result of
this population space having been incorrectly defined, rather than the result of how pre-data
knowledge about the distribution $p_x(y)$ has been expressed through the model $\mathcal{M}$.

As a result, it can be argued that there is only really one way in which the population space model
$\mathcal{M}$ or our expression of pre-data knowledge about the distribution $p_x(y)$ through this
model can be considered to be wrong or invalid, and that is if, on having more time for reflection,
we would define this model to be different from how it is currently defined, or express pre-data
knowledge about the distribution $p_x(y)$ through this model differently to how this pre-data
knowledge is currently expressed.
Of course, a decision to reflect more deeply on how pre-data knowledge about the true sampling
distribution\hspace{0.05em} $p_x(y)$ is best expressed through a population space model
$\mathcal{M}$ may be provoked by observing the data set $y$.
However, it should be clear that not carrying out this assessment of pre-data knowledge adequately
when we originally had the opportunity to do so may be justifiably classed as being a mistake, and
we would wish to discover this mistake at any moment in time, not necessarily when we have just
observed the data set $y$.
Also, even though, in the case being discussed, it is only on observing the data set $y$ that we
decide to re-assess how pre-data knowledge about the sampling distribution $p_x(y)$ is expressed
through the model $\mathcal{M}$, the result of this re-assessment may be that this expression of
pre-data knowledge about $p_x(y)$ does not change or, in a certain sense, moves more into conflict
with the data $y$.
Therefore, it should be clear that in this case, the population space model $\mathcal{M}$ or the
expression of pre-data knowledge about the true sampling distribution $p_x(y)$ through this model
is not being directly brought into question by the data set $y$.

\vspace{3ex}
\noindent
\section{Conclusion}
\label{sec9}

The key elements of the approach to inference that has been put forward in the present paper are as
follows:

\vspace{2ex}
\noindent
1) The explicit or implicit partitioning of an open, and therefore non-measurable, population space
into a finite, countable or uncountable number of subsets such that a probability distribution can
be placed over a variable that is defined so that it simply indicates which one of these subsets
contains the true sampling distribution $p_x(y)$.

\vspace{2ex}
\noindent
2) The possibility of being able to form a set of representative sampling distributions that belong
to an elegant parametric structure by selecting a sampling distribution from each of the subsets
that constitute the aforementioned type of partition that adequately represents all the other
sampling distributions in the subset concerned.

\vspace{2ex}
\noindent
3) The use of sampling distributions that do not necessarily belong to the same parametric family
of distributions to represent the type of subsets in question of the population space of interest
to allow for the fact that it may be much easier to express pre-data knowledge about the sampling
distribution $p_x(y)$ over a number of simple structures rather than one complicated structure.

\vspace{2ex}
\noindent
4) The lack of any fundamental role for P values in the modelling process of any given data set due
to the fact that it may be argued, as has been done in the present paper, that using a P value to
measure the absolute compatibility of a sampling distribution, or a parametric family of sampling
distributions, with an observed data set is neither meaningful nor useful.

\vspace{2ex}
\noindent
5) The conceptualisation of a statistical model as being a model of a population space rather than
a model of a sampling distribution with the result that it is arguably not possible for a
statistical model as conceived in the present paper or our expression of pre-data knowledge about
the sampling distribution $p_x(y)$ through such a model to be directly brought into question by the
data $y$.

\vspace{2ex}
\noindent
6) The use of the method of organic fiducial inference that is outlined in the Appendix of the
present paper (which can be found after the bibliography) in the case where the population space is
closed, or alternatively, if this closed population space is very large in terms of the number of
distributional families it contains or the number of parameters on which it depends, the use of an
approach that models this space in the same way in which it has been shown that an open population
space may be modelled.

\vspace{5.5ex}
\pdfbookmark[0]{Bibliography}{toc1}
\noindent
\textbf{Bibliography}

\begin{description}

\setlength{\itemsep}{1ex}

\vspace{0.5ex}
\item[] Bowater, R. J. (2018).\ Multivariate subjective fiducial inference.\ \textit{arXiv.org
(Cornell University), Statistics}, arXiv:1804.09804.

\item[] Bowater, R. J. (2019).\ Organic fiducial inference.\ \textit{arXiv.org (Cornell
University), Sta\-tis\-tics}, arXiv:1901.08589.

\item[] Bowater, R. J. (2021).\ A revision to the theory of organic fiducial inference.\
\textit{arXiv.org (Cornell University), Statistics}, arXiv:2111.09279.

\item[] Bowater, R. J. (2022).\ Sharp hypotheses and organic fiducial inference.\ \textit{arXiv.org
(Cornell University), Statistics}, arXiv:2207.08882.

\pagebreak
\item[] Bowater, R. J. (2023).\ The fiducial-Bayes fusion:\ a general theory of statistical
inference.\ \textit{arXiv.org (Cornell University), Statistics}, arXiv:2310.01533.

\item[] Clyde, M. A. (1999).\ Bayesian model averaging and model search strategies (with
discussion).\ In \textit{Bayesian Statistics 6}, Eds.\ J. M. Bernardo, J. O. Berger, A. P. Dawid
and A. F. M. Smith, Oxford University Press, Oxford, pp.\ 157--185.

\item[] Davies, L. (2014).\ \textit{Data Analysis and Approximate Models:\ Model Choice,
Location-Scale, Analysis of Variance, Nonparametric Regression and Image Analysis}, CRC Press,
Boca Raton.

\item[] Davies, L. (2018).\ On P-values.\ \textit{Statistica Sinica}, \textbf{28}, 2823--2840.

\item[] Draper, D. (1995).\ Assessment and propagation of model uncertainty (with discussion).\
\textit{Journal of the Royal Statistical Society, Series B}, \textbf{57}, 45--97.

\item[] Draper, D. (1999).\ Model uncertainty, yes, discrete model averaging, maybe.\ Comment on
`Bayesian model averaging:\ a tutorial' by J. A. Hoeting, D. Madigan, A. E. Raftery and C. T.
Volinsky.\ \textit{Statistical Science}, \textbf{14}, 405--409.

\item[] Greenland, S. (2023).\ Divergence versus decision P-values:\ a distinction worth making in
theory and keeping in practice: or, how divergence P-values measure evidence even when decision
P-values do not.\ \textit{Scandinavian Journal of Statistics}, \textbf{50}, 54--88.

\item[] Hoeting, J. A., Madigan, D., Raftery, A. E. and Volinsky, C. T. (1999).\ Bayesian model
averaging:\ a tutorial (with discussion).\ \textit{Statistical Science}, \textbf{14}, 382--417.

\item[] O'Hagan, A. (1995).\ Fractional Bayes factors for model comparison (with discussion).\
\textit{Journal of the Royal Statistical Society, Series B}, \textbf{57}, 99--138.

\item[] Rubin, D. B. (1984).\ Bayesianly justifiable and relevant frequency calculations for the
applied statistician.\ \textit{Annals of Statistics}, \textbf{12}, 1151--1172.

\item[] Shafer, G. (1982).\ Lindley's paradox (with discussion).\ \textit{Journal of the American
Statistical Association}, \textbf{77}, 325--351.

\end{description}

\pagebreak
\pdfbookmark[0]{Appendix. Organic fiducial inference for model comparison}{toc2}
\noindent
\textbf{Appendix}

\vspace{2.75ex}
\noindent
\textbf{Organic fiducial inference for model comparison}

\vspace{1.5ex}
\noindent
In this appendix, we will address the problem of how to determine an appropriate post-data density
for a population quantity of interest $Q$, i.e.\ a post-data density of the type $\pi(Q\,|\,y)$, in
the case where the sampling distribution $p_x(y)$ that generated the observed data set $y$ has to
be a member of a distributional family $F_i$ that, although being unknown, must belong to a given
closed population space $\mathcal{P}_C = \{F_1, F_2, \ldots, F_c\}$. In particular, this problem of
inference will be addressed by naturally extending a method of inference put forward in
Bowater~(2022) that belongs to an inferential framework known as the fiducial-Bayes fusion, which
was outlined in Bowater~(2023).

Let us begin by making an analogy between, on the one hand, our uncertainty, before the data set
$y$ was observed, about both the values that will make up this data set and the form of the
sampling distribution $p_x(y)$ that will generate this data set, and on the other hand, our
uncertainty about the random outcome of a well-understood physical experiment.
More specifically, we will assume that the physical experiment proceeds by, first, selecting at
random a sampling distribution from among the $c$ distributions contained in the set
$\{F_1[\theta_0^{(1)}], F_2[\theta_0^{(2)}], \ldots, F_c[\theta_0^{(c)}]\}$, and then drawing at
random the data set $y$ from this sampling distribution, where, for $i=1,2,\ldots,c$, the
distribution $F_i[\theta_0^{(i)}]$ is the sampling distribution belonging to the distributional
family $F_i$ that corresponds to the parameters $\theta^{(i)}$ of this family being equal to given
values $\theta_0^{(i)}$.
The key detail that establishes this analogy as generally being an acceptable one to make is that
it will be assumed that the values of the vectors in the set $\{\theta_0^{(1)}, \theta_0^{(2)},
\ldots, \theta_0^{(c)}\}$ are unknown to us, and in fact, it will be assumed that, in general, the
extent of what we know about the values of the vectors in this set does not enable us to identify a
positive measure subset of the population space $\mathcal{P}_C = \{F_1, F_2, \ldots, F_c\}$ that we
can be sure does not contain the sampling distribution that will generate the data set $y$.
To clarify, in conducting the physical experiment of interest, the joint probability density
function of the sampling distribution $F_j[\theta^{(j)}]$ that will be selected by this experiment
and the resulting data set $y$ would be given by:
\vspace{0.5ex}
\[
\pi(F_j[\theta^{(j)}],y) = \pi(F_j[\theta^{(j)}])\pi(y\,|\,F_j[\theta^{(j)}])
\]
where $j$ may take any value in the set $\{1,2,\ldots,c\}$, the function
$\pi(y\,|\,F_j[\theta^{(j)}])$ is the joint density function of the data $y$ given that the
sampling distribution of this data is $F_j[\theta^{(j)}]$ and where:
\vspace{1.5ex}
\[
\pi(F_j[\theta^{(j)}]) = \left\{
\begin{array}{ll}
\pi(F_j[\theta_0^{(j)}]) & \mbox{if $F_j[\theta^{(j)}] = F_j[\theta_0^{(j)}]$}\\[0.5ex]
0 & \mbox{otherwise}
\end{array}
\right.
\vspace{1ex}
\]
in which $\pi(F_j[\theta_0^{(j)}])$ is the probability that the sampling distribution
$F_j[\theta_0^{(j)}]$ is selected by the experiment.

We may observe that if the values of the vectors in the set $\{\theta_0^{(1)}, \theta_0^{(2)},
\ldots, \theta_0^{(c)}\}$ were known to us, then, on using our pre-data knowledge about the true
sampling distribution $p_x(y)$ to assign values to the prior probabilities of this distribution
belonging to each of the distributional families $\{F_1,F_2,\ldots,F_c\}$, i.e.\ the prior
probabilities $\{\pi(F_i):i=1,2,\ldots,c\}$ or the prior probabilities
$\{\pi(F_i[\theta_0^{(i)}]):i=1,2,\ldots,c\}$, the posterior probabilities of the distribution
$p_x(y)$ belonging to each of these distributional families would, according to the Bayesian
paradigm, be given by the expression:
\vspace{0.5ex}
\begin{equation}
\label{equ6}
\pi(F_j\,|\,y) = \pi(F_j[\theta_0^{(j)}]\,|\,y) =
\frac{\pi(F_j)\pi(y\,|\,F_j[\theta_0^{(j)}])}{\sum_{i=1}^{c}
\pi(F_i)\pi(y\,|\,F_i[\theta_0^{(i)}])}\ \ \ \mbox{for $j \in \{1,2,\ldots,c\}$}
\vspace{1ex}
\end{equation}
Also, let us point out that in the scenario of interest, i.e.\ in the case where the values of the
vectors in the set $\{\theta_0^{(1)},\theta_0^{(2)},\ldots,\theta_0^{(c)}\}$ are unknown to us, the
only genuine stumbling block we would face in trying to use the same expression to determine the
posterior probabilities $\{\pi(F_j\,|\,y):j=1,2,\ldots,c\}$ is that the set of sampling densities
$\{\pi(y\,|\,F_i[\theta_0^{(i)}]):i=1,2,\ldots,c\}$ will be unknown to us.
To get around this difficulty, we may apply a similar strategy to one put forward in Bowater~(2022)
by trying to estimate each of the sampling densities in the set
$\{\pi(y\,|\,F_i[\theta_0^{(i)}]):i=1,2,\ldots,c\}$ using an appropriate post-data predictive
density of a future data set $y'$ that is derived using organic fiducial inference, as outlined in
Bowater~(2019, 2021), under the condition that the sampling distribution $p_x(y)$ belongs to the
distributional family $F_i$ concerned. More precisely, we may define each of these post-data or
fiducial predictive densities as\vspace{0.5ex} follows:
\begin{equation}
\label{equ7}
f(y'\,|\,F_i,y) = \int_{\mbox{\footnotesize $\Theta^{(i)}$}} \pi(y'\,|\,F_i[\theta^{(i)}])
f(\theta^{(i)}\,|\,F_i,y)d\theta^{(i)}\ \ \ \mbox{for $i \in \{1,2,\ldots,c\}$}
\vspace{0.5ex}
\end{equation}
where:\\[1ex]
i) The function $f(\theta^{(i)}\,|\,F_i,y)$ is a joint fiducial density of the set of parameters
$\theta^{(i)}$ determined under the condition that the sampling distribution $p_x(y)$ belongs to
the distributional family $F_i$\hspace{0.1em}. It will be assumed that this fiducial density is
derived using the methodology outlined in Bowater~(2019, 2021) on the basis of whatever is
precisely assumed regarding our pre-data knowledge about the parameters $\theta^{(i)}$.\\[1ex]
ii) The future data set $y'$ is the same size as the observed data set $y$, and if applicable, is
derived on the basis of the same values for the conditioning variables $x$ that were used to derive
the data set $y$.\\[1ex]
iii) As earlier, $\Theta^{(i)}$ denotes the set of all possible values of the parameters
$\theta^{(i)}$.

\vspace{1ex}
On substituting the fixed but unknown sampling densities
$\{\pi(y\,|\,F_i[\theta_0^{(i)}]):i=1,2,\linebreak\ldots,c\}$ that appear in equation~(\ref{equ6})
by the estimates of these densities that were just put forward, i.e.\ the fiducial predictive
densities $\{f(y'\,|\,F_i,y):i=1,2,\ldots,c\}$ given by equation~(\ref{equ7}), the posterior
probabilities $\{\pi(F_j\,|\,y):j=1,2,\ldots,c\}$ as expressed in equation~(\ref{equ6}), which were
derived purely under the Bayesian paradigm, become simply justifiable post-data probabilities of
the sampling distribution $p_x(y)$ belonging to each of the distributional families
$\{F_1,F_2,\ldots,F_c\}$ that are defined as:
\vspace{1ex}
\begin{equation}
\label{equ8}
\bm\tilde{\pi}(F_j\,|\,y) = \frac{\pi(F_j)f(y'\hspace{-0.05em}=y\,|\,F_j,y)}{\sum_{i=1}^{c}
\pi(F_i)\pi(y'\hspace{-0.05em}=y\,|\,F_i,y)}\ \ \ \mbox{for $j \in \{1,2,\ldots,c\}$}
\vspace{1ex}
\end{equation}
Questions that some may choose to raise concerning the validity of the strategy that has been
applied to derive these post-data probabilities are best considered in the context of the longer
exposition of this type of strategy that is presented in Bowater~(2022).
Therefore, instead of addressing these questions here, we will restrict the current analysis to
simply clarifying how the problem of inference that we are interested in may be resolved using
ideas related to the strategy under discussion.

In this respect, let us conclude by noting that on substituting the generic post-data probabilities
$\{\pi(F_i\,|\,y):i=1,2,\ldots,c\}$ that appear in the definition of the post-data density
$\pi(Q\,|\,y)$ in equation~(\ref{equ2}) by the more specific post-data probabilities
$\{\bm\tilde{\pi}(F_i\,|\,y):i=1,2,\ldots,c\}$ as given in equation~(\ref{equ8}), it becomes clear
that, in the example being studied, the overall post-data density of a given quantity of interest
$Q$ is naturally defined as follows:
\vspace{1.5ex}
\[
\pi(Q\,|\,y) = \sum_{i=1}^{c} \int_{\mbox{\footnotesize $\Theta^{(i)}$}}
\pi(Q\,|\,F_i[\theta^{(i)}])f(\theta^{(i)}\,|\,F_i,y)\bm\tilde{\pi}(F_i\,|\,y) d\theta^{(i)}
\vspace{1ex}
\]
where, just to clarify, $\pi(Q\,|\,F_i[\theta^{(i)}])$ is the sampling density of the quantity $Q$
given that $F_i[\theta^{(i)}]$ is the sampling distribution of the data set $y$ and all other
notation is the same as has been used in this appendix.

\end{document}